\documentclass[nofootinbib,secnumarabic]{revtex4}
\usepackage{epsf,amsmath,amssymb,graphicx,dcolumn}
\usepackage{scalefnt}
\usepackage{hyperref}
\usepackage{cleveref,etoolbox,color,soul}
\usepackage{listings,booktabs}
\usepackage[utf8]{inputenc}
\usepackage{epsfig,psfrag}

\numberwithin{equation}{section}
\makeatletter       

\renewcommand{\p@subsection}{}

\makeatother

\newenvironment{Eqnarray}%
     {\arraycolsep 0.14em\begin{eqnarray}}{\end{eqnarray}}
\newcommand{\ba}{\begin{Eqnarray}}
\newcommand{\ea}{\end{Eqnarray}}
\newcommand{\be}{\begin{equation}}
\newcommand{\ee}{\end{equation}}

\begin{document}

\title{Stability of neutral minima against charge breaking in the Higgs triplet model}

\author{P.M.~Ferreira}
\email[E-mail: ]{pmmferreira@fc.ul.pt}
\affiliation{Instituto Superior de Engenharia de Lisboa, Instituto Polit\'ecnico de Lisboa
  1959-007 Lisboa, Portugal}
\affiliation{Centro de F\'{\i}sica Te\'{o}rica e Computacional,
Faculdade de Ci\^{e}ncias, Universidade de Lisboa, Campo Grande, Edif\'{\i}cio C8
1749-016 Lisboa, Portugal}
\author{B.L.~Gon\c{c}alves}
\email[E-mail: ]{bernardo.lopes.goncalves@tecnico.ulisboa.pt}
\affiliation{Departamento de Física and CFTP, Instituto Superior T\'ecnico, Universidade de Lisboa, 1049-001 Lisboa, Portugal}

\date{\today}

\begin{abstract}
We analyse the possibility of charge breaking minima developing in the Higgs triplet model, and under
what conditions they are deeper than charge-preserving ones. Analytical expressions relating the
depth of minima of different types are deduced. A global symmetry of the model leads to increased
stability for charge-preserving vacua. However, if that symmetry is broken by a soft term, deeper
charge-breaking minima may occur more easily. We identify the vev configurations most likely
to produce charge breaking minima.
\end{abstract}

\maketitle

\section{Introduction}
\label{sec:int}

One of the fundamental hypotheses of the Standard Model (SM) of particle physics is the
Higgs mechanism, through which elementary particles acquire their mass. These masses arise
from the spontaneous breaking of gauge symmetries, when a scalar doublet acquires a non-zero
vacuum expectation value (vev). This mechanism
implies the existence of an elementary spin-0 particle, the Higgs boson, finally discovered
in 2012 by the LHC collaborations~\cite{Aad:2012tfa,Chatrchyan:2012xdj}. Further
measurements of the properties of this particle
(see, for instance,~\cite{Aad:2015zhl,Khachatryan:2016vau})
show that it behaves in a very similar manner to the SM Higgs particle, but current precision
on the couplings of this scalar still leave a lot of room for theories with extended scalar
sectors. One of the simplest such models is the two-Higgs doublet model (2HDM), proposed
by Lee in 1973~\cite{Lee:1973iz}, as a means of introducing a new source of CP violation
in the model via spontaneous symmetry breaking. The 2HDM has a richer scalar spectrum than
the SM, including a heavier CP-even scalar, a pseudoscalar and a charged one, the possibility
of tree-level FCNC mediated by scalars and a more complex vacuum structure, including a
dark matter phase (see~\cite{Branco:2011iw} for a 2HDM review). Other theories
with extended scalar sectors popular in the literature include the doublet-singlet
model, where a $SU(2)\times U(1)$ gauge singlet is added to the model. This singlet
can either be real~\cite{McDonald:1993ex} or complex~\cite{Barger:2008jx}, and such
models are used to provide explanations for the dark matter relic abundance and
the first order electroweak baryogenesis phase transition.

The Higgs-triplet model (HTM)~\cite{Mohapatra:1979ia,Konetschny:1977bn,Magg:1980ut,
Cheng:1980qt,Schechter:1980gr,Lazarides:1980nt} is another possibility -- in addition to the
usual Higgs doublet, a
scalar triplet, with hypercharge $Y = 2$, is included. The scalar spectrum becomes much richer --
in the most common version of the model it includes two CP-even scalars, $h$ and $H$, a
pseudoscalar $A$, a charged scalar $H^\pm$ and a doubly charged one, $H^{\pm\pm}$. One of the
motivations of the model is the possibility of explaining the smallness of the neutrino masses
via a type-II seesaw mechanism. The Higgs-triplet model can also accommodate dark matter candidates,
and boasts a rich phenomenology. The presence of doubly charged scalars, in particular,
provides an interesting search channel for collider searches.

There is an extensive body of work on the Higgs-triplet model (see for
instance~\cite{Arhrib:2011uy,Arhrib:2011vc,Kanemura:2012rj,Aoki:2012yt,Aoki:2012jj,Xu:2016klg}
for recent works in this model), both on its theory underpinnings
and its phenomenological consequences. In this paper we will perform an in-depth analysis of
the vacuum structure of the model, using techniques developed to study the 2HDM. In particular,
we are interested in the possibility of charge-breaking (CB) vacua developing-- since the model
includes charged scalars, solutions of the minimization conditions of the potential which include
vevs possessing electrical charge are {\em a priori} possible. Since such solutions
would implicate a non-zero photon mass, the combinations of potential parameters which generate
them should be excluded. This CB vacuum analysis therefore provides us a tool which allows
us to limit the model's allowed parameter space, increasing its predictive power. Following
the 2HDM example, we will deduce analytical expressions which allow us to compare the depth of the scalar
potential at extrema which break different symmetries. The authors of~\cite{Arhrib:2011uy,Arhrib:2011vc}
performed a partial and qualitative analysis of CB in the HTM, we now propose to go further
in this vacuum analysis. In ref.~\cite{Xu:2016klg} analytical expressions relating the depths of
the HTM vacua were obtained, but in a different formalism than we will be employing. The author
of~\cite{Xu:2016klg} analysed in great (and exact) detail the HTM model with a specific global
symmetry intact, obtaining
approximate expressions for the relative depth of the potential in different extrema
for the case where that symmetry is softly broken. In the current work our expressions will be exact,
allowing a soft breaking parameter to have arbitrary size.

We will see that the doublet plays a special role in the vacuum stability picture of the model. Indeed,
if the scalar potential has a specific global symmetry, the neutral vacua of the model are completely
stable against charge breaking vacua if such vacua include vevs for both the doublet and the triplet; if
however the doublet has no vevs, deeper CB minima may exist below neutral ones. As such, unlike the 2HDM
case, the HTM neutral vacua are {\em not} guaranteed to be stable against charge breaking. If the global
symmetry mentioned above is softly broken by a given dimension-3 term, the picture of stability we have just
described further changes, and other deeper CB minima, and tunneling to them, become possible.

This paper is organized as follows:
we will describe the model in section~\ref{sec:mod}, with emphasis on the scalar sector and
possible vacua. In section~\ref{sec:nsb} we will discuss the vacuum structure of the potential without
the soft-breaking term, developing a field bilinear formalism for this model analogous to the
one employed for the 2HDM. Then, in section~\ref{sec:sb} we will allow for the presence of a
global symmetry soft breaking term, and show how it changes the stability of neutral
vacua with respect to charge breaking vacua. In section~\ref{sec:sher} we will study the situation where
 the doublet has no vev and the triplet possesses charge breaking vevs, and show how this changes the
 stability status of the potential without soft breaking term. We will perform a numerical study of the
model, investigating the regions of parameter space where deeper CB vacua might occur, in
section~\ref{sec:num}, and draw our conclusions in section~\ref{sec:con}. The appendices will
include a long list of analytical expressions mentioned in the main text, detailing the
differences in depths of the potential at different stationary points with complex vevs.

\section{The Higgs-triplet model}
\label{sec:mod}

The Higgs-triplet model (HTM) is an extension of the scalar sector of the Standard Model (SM). The
gauge symmetries and fermionic content are (usually) the same in both models, but the HTM contains
a larger scalar sector, wherein the hypercharge 1 Higgs doublet of the SM, $\Phi$, is complemented
by a hypercharge 2 triplet, $\Delta$. These fields may be written as
\be
\Phi\,=\,\left(
\begin{array}{c} \phi^+ \\ \phi^0\end{array}
\right)\;\;\; , \;\;\;
\Delta = \left( \begin{array}{cc} \Delta^+/\sqrt{2} & \Delta^{++} \\
\Delta^0 & - \Delta^+/\sqrt{2} \end{array} \right)
\ee
where all $\phi^x$, $\Delta^y$ are complex fields, and we are using a $SU(2)$ matrix representation
for the triplet $\Delta$. Notice the occurrence of doubly-charged scalars in the theory, a consequence
of the hypercharge assignment of the triplet field. The most general $SU(2)\times U(1)$ scalar potential
involving these two fields is then given by
\begin{align}
V &= m^2 \Phi^\dagger\Phi\,+\,M^2 \mbox{Tr}(\Delta^\dagger\Delta)\,+\,\mu\left(
\Phi^T \mbox{i}\tau_2 \Delta^\dagger \Phi\,+\,\mbox{h.c.} \right) \nonumber \\
 &+ \lambda_1 (\Phi^\dagger\Phi)^2 \,+\, \lambda_2 \left[\mbox{Tr}(\Delta^\dagger\Delta)\right]^2
 \,+\,\lambda_3 \mbox{Tr}\left[(\Delta^\dagger\Delta)^2\right]\,+\,
 \lambda_4 (\Phi^\dagger\Phi) \mbox{Tr}(\Delta^\dagger\Delta)\,+\,
 \lambda_5 \Phi^\dagger \Delta \Delta^\dagger \Phi\, ,
 \label{eq:V}
\end{align}
with all parameters in the potential being real, and $h.c.$ standing for ``hermitian
conjugate". So that the model is bounded from below -- and therefore possesses a stable
global minimum -- the quartic couplings $\lambda_{1,\dots 5}$ must obey the
following necessary and sufficient conditions~\cite{Arhrib:2011uy}:
\begin{align}
 & \lambda_1 \,>\, 0 \;\;\; , \;\;\;
\lambda_2 \,+\, \mbox{min} \left(\lambda_3 \,,\,\dfrac{1}{2} \lambda_3\right) \,>\, 0 \, ,\nonumber  \\
& \lambda_4 \,+\, \mbox{min} \left(0\,,\, \lambda_3\right)\,+\,
2 \mbox{min} \left[\sqrt{\lambda_1 (\lambda_2 + \lambda_3)}\,,\,
\sqrt{\lambda_1 (\lambda_2 + \lambda_3/2)}\right] \,>\, 0 \,.
\label{eq:bfb}
\end{align}
In ref.~\cite{Arhrib:2011uy} bounds on the quartic couplings of the potential so that
the theory preserves unitarity were also presented.

Notice now the term cubic in the fields with coefficient $\mu$: it can be removed by
imposing on the potential, for instance, a global $U(1)$ symmetry of the form $\Phi\rightarrow
e^{i\theta} \Phi$, with $\theta$ an arbitrary real number~\footnote{In fact, even a discrete
$Z_2$ symmetry of the form $\Delta\rightarrow -\Delta$ would suffice to eliminate $\mu$ --
but one would be left with a potential which indeed possessed a global continuous symmetry.
This is an example of ``accidental" continuous symmetries arising from the imposition
of discrete ones, a well-known occurrence in the 2HDM and 3HDM~\cite{Ferreira:2008zy}.}.
Therefore, the $\mu$ term is a soft breaking of this global symmetry.
The theory without the soft breaking term, with the global symmetry intact,
 is phenomenologically interesting, since it allows for dark
matter particles. On the other hand, softly breaking this continuous global symmetry
is also of interest, since it can be used to help generate neutrino masses via the
seesaw mechanism. Both theories -- with or without the soft breaking term -- are therefore
relevant, and we will study their vacuum structure separately.

The Higgs-triplet model, of course, also includes fermions, and the scalar-fermion
interactions are contained in the Yukawa lagrangian.
The quarks, due to the hypercharge assignment of
all fields, do not interact at all with the scalar triplet. In the lepton sector
other possibilities arise due to the presence of terms such as
 $L\Delta L$, with $L$ being lepton left doublets. A seesaw mechanism can also be introduced to generate
 masses for neutrinos, but we will not be studying such matters in this work.
All that we then need for the current work is to
remember that the Yukawa lagrangian concerning quarks in this
model is identical to the SM's, and therefore all of those fermion masses will be proportional
to the doublet $\Phi$'s vev.
Therefore, any minimum where the doublet is vevless would be unphysical, since the quarks
would be massless.

\subsection{The neutral vacua}

The Higgs-triplet model has three possible minima wherein the vevs are neutral, and thus
electric charge conservation (and indeed all of electromagnetism) holds. However, they are
not identical, and indeed yield very different phenomenologies. For now we will consider
only real vevs, and in section~\ref{sec:cp} we will discuss the possibility of neutral
complex vevs. We call these possible minima with neutral real vevs {\em Normal minima},
and there are three possibilities:
\begin{itemize}
\item The $N1$ stationary point, where both the doublet and triplet have neutral vevs,
\be
\langle \Phi\rangle_{N1} = \frac{1}{\sqrt{2}}\,\left(
\begin{array}{c} 0 \\ v_\Phi \end{array}
\right)\;\;\; , \;\;\; \langle \Delta \rangle_{N1} = \frac{1}{\sqrt{2}}\,
\left( \begin{array}{cc} 0 & 0 \\
v_\Delta & 0 \end{array}  \right)\, .
\label{eq:N1}
\ee
Both vevs contribute to the gauge boson masses, and in order to have the correct
electroweak symmetry breaking one would need to have $v_\Phi^2 + 2v^2_\Delta \simeq$
(246 GeV)$^2$~\footnote{In fact the triplet's vev contributes differently to the W and Z masses
in the HTM -- one has $m^2_W = g^2 (v_\Phi^2 + 2v^2_\Delta)/4$ and
$m^2_Z = (g^2 + {g^\prime}^2) (v_\Phi^2 + 4 v^2_\Delta)/4$. Thus in this model the tree-level prediction
for the electroweak precision constraint parameter $\rho$ is not equal to 1, unlike models with an
arbitrary number of doublets. This then forces the triplet vev to be limited in magnitude,
typically no more than 8 GeV.}.
This extremum can occur whether the soft breaking $\mu$ term is present or not.
Defining
\be
M_{\Delta}^{2}\,\equiv\,\frac{v_{\Phi}^{2}\,\mu}{\sqrt{2}\,v_{\Delta}}\,,
\label{eq:MD}
\ee
the pseudoscalar, singly charged and doubly charged scalar masses are given by
\begin{align}
  m^2_{A} & = M_{\Delta}^2 \left( 1+\dfrac{4v_{\Delta}^{2}}{v_{\Phi}^{2}} \right)
  \label{eq:mA1}\\
  m^2_+ & = \left( M_{\Delta}^2 - \dfrac{\lambda_{5}}{4}v_{\Phi}^{2} \right) \left(
  1+\dfrac{2v_{\Delta}^{2}}{v^{2}_{\Phi}} \right)
  \label{eq:mch1}\\
  m^2_{++} &= M_{\Delta}^{2}-v_{\Delta}^{2}\lambda_{3}-\dfrac{\lambda_{5}}{2}v_{\Phi}^{2}\,.
  \label{eq:mchch1}
\end{align}
We therefore see that if the soft breaking $\mu$ term is not present we will have $M_{\Delta} = 0$
and consequently $m_A = 0$ -- the triplet vev spontaneously breaks a global continuous
symmetry and the theory develops a massless axion. As for the CP-even scalars $h$ and $H$,
their masses will be the eigenvalues of the $2\times 2$ matrix
\be
[m^2_{h,H}]\,=\,\left(\begin{array}{cc}
2\lambda_1 v_\phi^2 & -\frac{2 v_\Delta}{v_\phi}\,M_{\Delta}^2 +
 (\lambda_4 + \lambda_5)v_\phi v_\Delta \\
 -\frac{2 v_\Delta}{v_\phi}\,M_{\Delta}^2 +  (\lambda_4 + \lambda_5)v_\phi v_\Delta &
 M_{\Delta}^2 + 2 (\lambda_2 + \lambda_3) v^2_\Delta
\end{array}\right)\,.
\ee
\item The $N2$ stationary point, where only the doublet has a vev,
\be
\langle \Phi\rangle_{N2} = \frac{1}{\sqrt{2}}\,\left(
\begin{array}{c} 0 \\ v \end{array}
\right)\;\;\; , \;\;\; \langle \Delta \rangle_{N2} = \frac{1}{\sqrt{2}}\,
\left( \begin{array}{cc} 0 & 0 \\
0 & 0 \end{array}  \right)\,,
\label{eq:N2}
\ee
where in this case, to obtain the correct electroweak symmetry breaking, one
must have $v_\phi = 246$ GeV.
Unlike $N1$, this extremum can only exist if $\mu = 0$ and no soft-breaking of the global
continuous symmetry occurs. The doublet vev provides a mass to gauge bosons and fermions.
The two neutral states emerging from the doublet ($H$ and $A$) will be degenerate and are
good dark matter candidates. The scalar masses are given by
\begin{align}
m^2_h &= \,2\lambda_1 v_\Phi^2 \;\; , \label{eq:mh2}\\
m^2_H = m^2_A &= \, M^2 + \frac{1}{2} (\lambda_4 + \lambda_5) v_\Phi^2
\label{eq:mH2}
\;\; , \\
m^2_+ &=\, M^2 + \frac{1}{4} (2 \lambda_4 + \lambda_5) v_\Phi^2
\label{eq:mch2}
\;\; , \\
m^2_{++} &=\, M^2 + \frac{1}{2} \lambda_4 v_\Phi^2
\label{eq:mchch2}
\end{align}
where the SM-like Higgs boson is the $h$ state.
\item The $N3$ stationary point, where only the triplet has a vev,
\be
\langle \Phi\rangle_{N3} = \frac{1}{\sqrt{2}}\,\left(
\begin{array}{c} 0 \\ 0 \end{array}
\right)\;\;\; , \;\;\; \langle \Delta \rangle_{N3} = \frac{1}{\sqrt{2}}\,
\left( \begin{array}{cc} 0 & 0 \\
v_\Delta & 0 \end{array}  \right)\,.
\label{eq:N3}
\ee
This extremum is clearly unphysical -- absence of a doublet vev means that all
quarks would be massless. Therefore, we will want to avoid this vacuum if possible.
Notice that $N3$ is a possible solution to the minimization conditions whether the
soft breaking term $\mu$ is present or not. Since the masses at $N3$ will not be required
for the stability analysis that follows, we will not present them.
\end{itemize}

The neutral minima of greater interest for the softly broken potential is clearly $N1$ --
in that case $N2$ cannot occur and an $N3$ minimum would imply massless quarks. On the
other hand, if the potential has a global continuous symmetry that is not softly broken
by the $\mu$ term, then it is $N2$ the neutral minimum that is relevant for particle
physics phenomenology -- $N1$ would imply a massless axion, and $N3$ is, once again,
unphysical.

\subsection{Spontaneous CP breaking?}
\label{sec:cp}

We have not considered complex neutral vevs in the previous section for a simple reason:
spontaneous CP breaking, where such vevs could arise, is not possible in the Higgs-triplet model.
The demonstration of this property is remarkably simple: first, consider that the $\mu$ parameter
can indeed always be rendered real, by performing a basis change on the doublet $\Phi$, for instance.
This means that if by any chance one were to consider a complex soft breaking parameter, $\mu = |\mu| e^{i \alpha}$,
one could redefine the doublet $\Phi$ such that $\Phi^\prime = e^{-i \alpha/2} \Phi$, thus
eliminating the phase $\alpha$ from the scalar potential -- and the theory, expressed in terms of
the doublet $\Phi^\prime$, would be exactly the same as before~\footnote{The phase $\alpha$ could also be
removed from the Yukawa sector by means of a phase redefinition of all right handed fermions,
for example.}.

Thus the writing of the potential in eq.~\eqref{eq:V} is indeed the most general potential,
naturally CP invariant under the symmetry $\Phi\rightarrow \Phi^*$ and $\Delta\rightarrow \Delta^*$.
A vacuum with spontaneous CP breaking would involve neutral complex vevs, and again via basis
redefinitions one can choose to have a single phase in, for instance, the triplet vev. The most
general possible CP-breaking vevs will therefore be given by
\be
\langle \Phi\rangle_{CP} = \frac{1}{\sqrt{2}}\,\left(
\begin{array}{c} 0 \\ v_1 \end{array}
\right)\;\;\; , \;\;\; \langle \Delta \rangle_{CP} = \frac{1}{\sqrt{2}}\,
\left( \begin{array}{cc} 0 & 0 \\
v_2 e^{i \theta} & 0 \end{array}  \right)
\ee
with real $v_1$ and $v_2$. Substituting these vevs into the potential of eq.~\eqref{eq:V}
we obtain
\be
V_{CP} \,=\, a \,+\, \mu b\,\cos\theta\,,
\ee
where the real coefficients $a$ and $b$ are functions of the parameters of the potential
and the magnitudes $v_1$ and $v_2$ -- all $\theta$ dependence is contained in the $\cos\theta$
term above. If there is no soft-breaking $\mu = 0$, the vev phase dependence in the potential
vanishes completely. With a softly broken potential the value of $\theta$ may be determined
by the minimization equations -- and clearly, $\partial V/\partial \theta = -\mu \,b \,\sin\theta = 0$
implies that any extremum of the potential will have $\theta = n\pi$, and thus no complex
phases between the vevs are possible.

\subsection{The charge breaking vacua}

For charge breaking (CB) to occur one or more vevs carrying electrical charge need to
appear as a result from spontaneous symmetry breaking. Such vevs would generate a non-zero
photon mass, in complete disagreement with the observed behaviour of electromagnetic phenomena.
Both the doublet and triplet fields have charged components, so there is a varied
assortment of possible CB extrema. In the main body of the paper we will only address
real vevs, but in the appendices we will show the results obtained for CB vevs with
imaginary parts -- such imaginary vevs do not bring any new features that cannot be
established by looking at real vevs alone. Before presenting the several CB vev patterns,
let us remember that one can always, via a suitable gauge choice, absorb three real scalar
component fields. Such a gauge choice will of course affect both the doublet and the triplet,
and we choose to be analogous to the SM unitary gauge -- wherein the doublet is reduced to a neutral,
real component. As such, in all field vevs dealt with in this paper, the doublet vev is always
real and neutral, without loss of generality. In some cases -- such as the absence of the
soft breaking term $\mu$ -- independent overall phase redefinitions of both the doublet and the triplet
are also possible, which leads to further simplifications of vevs, when complex phases are possible.

There are six possible real CB vev choices, cases $CB1$ to $CB6$:
\ba
\langle \Phi\rangle_{CB1} = \frac{1}{\sqrt{2}}\,\left(
\begin{array}{c} 0 \\ c_1 \end{array}
\right)\;\;\; &,& \;\;\; \langle \Delta \rangle_{CB1} = \frac{1}{\sqrt{2}}\,
\left( \begin{array}{cc} -c_3/\sqrt{2} & 0 \\
c_2 & c_3/\sqrt{2} \end{array}  \right)
\label{eq:CB1} \\
\langle \Phi\rangle_{CB2} = \frac{1}{\sqrt{2}}\,\left(
\begin{array}{c} 0 \\ c_1 \end{array}
\right)\;\;\; &,& \;\;\; \langle \Delta \rangle_{CB2} = \frac{1}{\sqrt{2}}\,
\left( \begin{array}{cc} 0 & c_3 \\
c_2 & 0 \end{array}  \right)
\label{eq:CB2} \\
\langle \Phi\rangle_{CB3} = \frac{1}{\sqrt{2}}\,\left(
\begin{array}{c} 0 \\ c_1 \end{array}
\right)\;\;\; &,& \;\;\; \langle \Delta \rangle_{CB3} = \frac{1}{\sqrt{2}}\,
\left( \begin{array}{cc} c_3/\sqrt{2} & c_4 \\
c_2 & -c_3/\sqrt{2} \end{array}  \right)
\label{eq:CB3} \\
\langle \Phi\rangle_{CB4} = \frac{1}{\sqrt{2}}\,\left(
\begin{array}{c} 0 \\ c_1 \end{array}
\right)\;\;\; &,& \;\;\; \langle \Delta \rangle_{CB4} = \frac{1}{\sqrt{2}}\,
\left( \begin{array}{cc} c_2/\sqrt{2} & 0 \\
0 & -c_2/\sqrt{2} \end{array}  \right)
\label{eq:CB4} \\
\langle \Phi\rangle_{CB5} = \frac{1}{\sqrt{2}}\,\left(
\begin{array}{c} 0 \\ c_1 \end{array}
\right)\;\;\; &,& \;\;\; \langle \Delta \rangle_{CB5} = \frac{1}{\sqrt{2}}\,
\left( \begin{array}{cc} c_2/\sqrt{2} & c_3 \\
0 & -c_2/\sqrt{2} \end{array}  \right)
\label{eq:CB5} \\
\langle \Phi\rangle_{CB6} = \frac{1}{\sqrt{2}}\,\left(
\begin{array}{c} 0 \\ c_1 \end{array}
\right)\;\;\; &,& \;\;\; \langle \Delta \rangle_{CB6} = \frac{1}{\sqrt{2}}\,
\left( \begin{array}{cc} 0 & c_2 \\
0 & 0 \end{array}  \right)
\label{eq:CB6}
\ea
The case $CB3$ is clearly the most generic real CB vev pattern possible for this model,
with vevs both neutral ($c_1$ and $c_2$), carrying a single charge ($c_3$) or a double
one ($c_4$). The remaining cases correspond to different possibilities, {\em a priori}
allowed by the minimisation conditions of the potential, where one or more of those
vevs are zero. Let us emphasise what is perhaps an obvious point: though for convenience
of notation we use $c_1$, $\dots$ $c_4$, to refer to all the CB vevs in the above six
cases, these quantities are {\em not} supposed to be equal for different CB extrema. For instance,
$c_2$ stands for a neutral vev in the $CB1$ case, a charged one for $CB4$ and doubly charged
one for $CB6$. For each of the CB cases considered, then, the value of the vevs $c_i$
will be determined by the minimisation of the potential, and depend on the model's
parameters. For the moment, until section~\ref{sec:sher}, we will always consider
$c_1 \neq 0$ -- the extrema considered will be such that the
doublet will always have a non-zero vev.

Also, given that we will be requiring {\em coexisting} normal and CB extrema, it will
occur in some cases that the minimisation conditions end up being impossible to solve
for some combinations of extrema, unless a specific combination of parameters of
the model is verified. We will draw attention to such cases when they occur.

\section{Potential without soft-breaking term}
\label{sec:nsb}

If a global continuous symmetry is imposed on the potential, the $\mu$ term in eq.~\eqref{eq:V}
is zero. There are no cubic terms in the potential, then, only quadratic, $V_2$, and quartic,
$V_4$, ones: $V = V_2 + V_4$. Any solution of the minimisation equations of the potential will
imply a simple
relation between the values of $V_2$ and $V_4$ at any stationary point (CP), to wit
\be
\mbox{At any stationary point:}\, \frac{\partial V}{\partial \varphi_i}\,=\,0
\,\Longrightarrow\,\sum_i \varphi_i\,\frac{\partial V}{\partial \varphi_i}\,=\,0\,\Longrightarrow
2\,V_2\,+\,4\,V_4\,=\,0\,.
\label{eq:homog}
\ee
This is a simple consequence of the potential being given by the sum of two homogenous functions
of the fields, $V_2$ a second degree homogenous function and $V_4$ a four degree one.
Therefore, the value of the potential at a given stationary point, $V_{SP}$, will be simply
\be
V_{SP}\;=\;\frac{1}{2}\,\left(V_2\right)_{SP}\;=\;-\,\left(V_4\right)_{SP}\,.
\label{eq:VSP}
\ee
We will be using this simplified expression for the value of the potential at an extremum
quite often.

Since the potential only has quadratic and quartic terms when $\mu = 0$, and we are
interested in comparing the value of the potential at different extrema, it is tempting to
attempt to use a bilinear formalism similar to the one employed for the 2HDM~
\cite{Velhinho:1994np,Ferreira:2004yd,Barroso:2005sm,Nishi:2006tg,Maniatis:2006fs,Ivanov:2006yq,
  Barroso:2007rr,Nishi:2007nh,Maniatis:2007vn,Ivanov:2007de,Maniatis:2007de,Nishi:2007dv,
  Maniatis:2009vp,Ferreira:2010hy}. Generalisations of this formalism have been used to
  study the vacuum structure of models other than the SM, for instance the 3HDM
\cite{Ivanov:2010ww,Ivanov:2014doa}, the complex singlet-doublet model
\cite{Ferreira:2016tcu} or the N2HDM~\cite{Ferreira:2019iqb}. We recall that in those works the bilinears
defined are always real gauge-invariant quantities, quadratic in the fields. Expressed as
a function of field bilinears, then, the scalar potential becomes a quadratic polynomial,
whose minimisation is simplified, and whose geometrical properties allow for an
in-depth analysis of potential symmetries and vacuum structure.

The major problem in attempting the formulation of a bilinear formalism for the Higgs-triplet
model are the $\lambda_3$ and $\lambda_5$ terms in the scalar potential of eq.~\eqref{eq:V},
which cannot obviously be written as the product of two terms quadratic in the fields. This would
seem to be an unsurmountable obstacle to a bilinear formulation but, in fact, can easily be overcomed.
For the study of the vacuum structure of the scalar potential, and the comparison of the value
of that potential at different extrema, we do not need the full field-dependent potential,
but rather only {\em the potential as a function of the vevs at the several stationary points}.
Let us then consider, as an example, the value of the potential at an $N1$ stationary point,
with vevs given by eqs.~\eqref{eq:N1},
\be
V_{N1} \,=\, \frac{1}{2} m^2 v_\Phi^2\,+\,\frac{1}{2} M^2 v_\Delta^2 \,+\,\frac{\lambda_1}{4} v_\Phi^4
\,+\, \frac{\lambda_2 + \lambda_3}{4} v_\Delta^4\,+\, \frac{\lambda_4 + \lambda_5}{4} v_\Phi^2 v_\Delta^2
\ee
and the potential at a $CB3$ stationary point, with vevs given by eqs.~\eqref{eq:CB3},
\ba
V_{CB3} &=& \frac{1}{2} \,m^2 c_1^2\,+\,\frac{1}{2} \,M^2 (c_2^2 +  c_3^2 +  c_4^2) \,+\,
\frac{\lambda_1}{4}  \,c_1^4 \,+\, \frac{\lambda_2 + \lambda_3}{4} \,(c_2^2 +  c_3^2 +  c_4^2)^2
\nonumber \\
& & -\, \frac{\lambda_3}{8} \, (c_3^2 +  2 c_2 c_4)^2\,+\,
\frac{\lambda_4}{4} \, c_1^2 (c_2^2 +  c_3^2 +  c_4^2) \,+\,
\frac{\lambda_5}{4} \, c_1^2 (2 c_2^2 +  c_3^2)\, .
\label{eq:VCB3}
\ea
Let us now define the vector $A$ and symmetric matrix $B$,
\be
A = \left(\begin{array}{c} m^2 \\ M^2 \\ 0 \\ 0 \\ 0
  \end{array}\right)\quad\quad ,\quad\quad
B = \left(\begin{array}{ccccc}
2 \lambda_1 & \lambda_4 + \lambda_5  & -\frac{1}{2} \lambda_5  & - \lambda_5 & 0 \\
\lambda_4 + \lambda_5 & 2(\lambda_2 + \lambda_3) & 0 & - 2 \lambda_3 & 0 \\
-\frac{1}{2} \lambda_5  & 0  & - \lambda_3 & 2 \lambda_3 & -  2 \lambda_3 \\
- \lambda_5  & - 2 \lambda_3 & 2 \lambda_3  & 4 \lambda_3 & 0 \\
 0 & 0 & - 2 \lambda_3 & 0 & 0
    \end{array}\right) \, ,
\label{eq:def}
\ee
and the five real quantities, $x_1 = |\Phi|^2$, $x_2 = \mbox{Tr}(\Delta^\dagger\Delta)$,
$x_3 = |\Delta^+|^2$, $x_4 = |\Delta^{++}|^2$ and $x_5 = |\Delta^0 \Delta^{--}|$. Then,
at each of the stationary points above, the entries of the vector $X$ (defined as $X^T =
(x_1\,,\,x_2\,,\,x_3\,,\,x_4\,,\,x_5)$) are given by
\be
X_{N1} = \frac{1}{2}\,\left(\begin{array}{c} v_\phi^2 \\ v_\Delta^2 \\ 0 \\ 0 \\ 0
  \end{array}\right)\quad\quad ,\quad\quad
X_{CB3} = \frac{1}{2}\,\left(\begin{array}{c} c_1^2 \\ c_2^2 +  c_3^2 +  c_4^2 \\ c_3^2
\\ c_4^2 \\ c_2 c_4
  \end{array}\right)\, ,
  \label{eq:xn1}
\ee
and simple algebra leads us to conclude that, for both of these stationary points,
the value of the potential can be written as
\be
V_{N1}\,=\, A^T X_{N1}\,+\,\frac{1}{2}\, X^T_{N1}\,B\,X_{N1}
\ee
and
\be
V_{CB3}\,=\, A^T X_{CB3}\,+\,\frac{1}{2}\, X^T_{CB3}\,B\,X_{CB3}\,.
\ee
In such expressions (and one obtains analogous ones for the remainder of the $N$ or $CB$ stationary
points) we recognise the structure of bilinears familiar from the 2HDM
case~\cite{Ferreira:2004yd,Barroso:2005sm}. Using eq.~\eqref{eq:VSP}, then, we find that at any
given stationary point the value of the potential may be expressed in very simple terms, as
\be
V_{SP}\;=\; \frac{1}{2}\,A^T X_{SP}\;=\; -\,X^T_{SP}\,B\,X_{SP}\,,
\label{eq:VSPb}
\ee
where $X_{SP}$ is the vector $X$ defined above evaluated at the vevs of the stationary point
under study.

Finally, let us define the vector $V^\prime$, the gradient of the potential with respect to the vector
$X$, which will play a crucial role in our stability analyses:
\be
V^\prime\;=\; \frac{\partial V}{\partial X^T}\;=\; A\,+\,B\,X\,.
\label{eq:Vl}
\ee

\subsection{Stability of minima of type $N2$ against charge breaking}

As we have explained above, for a Higgs-triplet model without soft breaking $\mu$ term the
$N2$ minimum is the most phenomenologically appealing. For such a stationary point, the
vectors $X$ and $V^\prime$ are given by
\be
X_{N2} \,=\, \frac{1}{2}\,\left(\begin{array}{c} v_\phi^2 \\ 0 \\ 0 \\ 0 \\ 0
  \end{array}\right)\quad\quad ,\quad\quad
V^\prime_{N2}\,=\,A\,+\,B\,X_{N2}\,=\,
\left(\begin{array}{c} 0 \\ m^2_{H,A} \\ -\frac{1}{4}\lambda_5\,v_\phi^2 \\
-\frac{1}{2}\lambda_5\,v_\phi^2 \\ 0
  \end{array}\right)\,,
  \label{eq:xn2vln2}
\ee
where the entries of $V^\prime_{N2}$ are a consequence of the minimisation conditions for
this extremum. The $m_{H,A}$ scalar mass at the $N2$ stationary point is
given by eq.~\eqref{eq:mH2}.

We must now verify the stability of a $N2$ minimum against the possibility of any deeper
charge breaking extrema. We will show how the demonstration is done for one specific example,
and leave the rest as an exercise for the readers. Let us suppose that for a given combination
of parameters of the potential there exist simultaneously stationary points of type $N2$ and $CB1$
-- {\em i.e.} the minimisation equations of the potential admit both types of solution.
For the vevs of a $CB1$ extremum, eq.~\eqref{eq:CB1}, and using the definitions of eqs.~\eqref{eq:def}
and~\eqref{eq:Vl}, we obtain the vectors $X_{CB1}$ and $V^\prime_{CB1}$,
\be
X_{CB1} \,=\, \frac{1}{2}\,\left(\begin{array}{c} c_1^2 \\ c_2^2 + c_3^2 \\ c_3^2 \\ 0 \\ 0
  \end{array}\right)\quad\quad ,\quad\quad
V^\prime_{CB1}\,=\,A\,+\,B\,X_{CB1}\,=\,-\, \left(\begin{array}{c} 0 \\ 0 \\ 0 \\ \frac{1}{2}\lambda_5 c_1^2
+ \lambda_3 c_2^2 \\ \lambda_3 c_3^2
  \end{array}\right)\,,
    \label{eq:xcb1vlcb1}
\ee
where once again the entries of $V^\prime$ are determined by the minimisation conditions
at this specific stationary point. In particular, the first entry being zero is a consequence of our having assumed
 $c_1\neq 0$, a point we will return to in section~\ref{sec:sher}.
 Attentive readers will notice that for both stationary
points considered the vectors $X$ and $V^\prime$ are orthogonal, $X^T V^\prime = 0$. This is
no coincidence, since $V^\prime$ is in fact the gradient of the potential $V$ along
the direction $X$. But let us now perform this internal product for vectors belonging
to {\em different} stationary points -- the result thereof is no longer zero in general,
and we obtain
\be
X^T_{CB1} V^\prime_{N2}\,=\,\frac{1}{2}\,\left[(c_2^2 + c_3^2) m^2_{H,A} \,-\,
\frac{1}{4}\lambda_5\,c_3^2\,v_\phi^2 \right]\,=\,X^T_{CB1} A\,+\,X^T_{CB1}\,B\,X_{N2}\,,
\ee
where we used both eqs.~\eqref{eq:xn2vln2} and~\eqref{eq:xcb1vlcb1}. Now, since for
an $N2$ stationary point $m^2_+ = m^2_{H,A} \,-\,\lambda_5\,v_\phi^2/4$ (check eqs.~\eqref{eq:mch2}
and~\eqref{eq:mchch2}) and from eq.~\eqref{eq:VSPb} we have $X^T_{CB1} A = 2 V_{CB1}$,
we can rewrite the equation above as
\be
V_{CB1}\,=\,\frac{1}{4}\,\left(c_2^2  m^2_{H,A} \,+\, c_3^2 m^2_{+} \right)
\,-\,\frac{1}{2}\,X^T_{CB1}\,B\,X_{N2}\,.
\label{eq:VCB1i}
\ee
Performing now the product of the other two vectors we have
\be
X^T_{N2} V^\prime_{CB1}\,=\,0 \,=\,X^T_{N2} A\,+\,X^T_{N2}\,B\,X_{CB1}\,,
\ee
whereupon we use eq.~\eqref{eq:VSPb} again to write $X^T_{N2} A = 2 V_{N2}$ and conclude
that
\be
V_{N2}\,=\,-\,\frac{1}{2}\,X^T_{N2}\,B\,X_{CB1}\,.
\label{eq:VN2i}
\ee
Notice now that, since the matrix $B$ is symmetric, the leftmost quantity in eqs.~\eqref{eq:VCB1i}
and~\eqref{eq:VN2i} is the same, so that when we subtract both equations it cancels and we obtain
\be
V_{CB1}\,-\,V_{N2}\;=\;\frac{1}{4}\,\left(c_2^2  m^2_{H,A} \,+\, c_3^2 m^2_{+} \right)\,.
\label{eq:VCB1N2}
\ee
This expression relates the depth of the potential at two stationary points of types $N2$ and $CB1$.
If $N2$ is a minimum, all of its squared masses will be positive and therefore the quantity in
left brackets above is perforce positive. Therefore, eq.~\eqref{eq:VCB1N2} implies that,
if $N2$ is a minimum, any $CB1$ stationary point that might exist is necessarily located
above $N2$ -- and therefore {\em $N2$ is stable against charge breaking vacua of type $CB1$}.

The method detailed above can be applied to any pairs of extrema. Following the same steps --
write down the $X$ and $V^\prime$ vectors at each stationary points; perform the product
of vectors from different stationary points; rewrite the results in terms of masses from
one of them and the values of the potential; eliminate the common term appearing in such products --
one can obtain the expressions relating the depth of $N2$ relative to any of the other five
CB extrema:
\ba
V_{CB2} - V_{N2} &=& \frac{1}{4} \,\left(c_2^2 \, m^2_{H,A} \,+\,c_3^2 \, m^2_{++}\right) \nonumber \\
& & \nonumber \\
V_{CB3} - V_{N2} &=& \frac{1}{4} \,\left(c_2^2 \, m^2_{H,A} \,+\,c_3^2 \, m^2_{+} \,+\,c_4^2 \, m^2_{++}\right) \nonumber \\
& & \nonumber \\
V_{CB4} - V_{N2} &=& \frac{1}{4} \, c_2^2 \, m_{+}^{2} \nonumber \\
& & \nonumber \\
V_{CB5} - V_{N2} &=& \frac{1}{4} \,\left(c_2^2 \, m^2_{+} \,+\,c_3^2 \, m^2_{++}\right) \nonumber \\
& & \nonumber \\
V_{CB6} - V_{N2} &=& \frac{1}{4} \, c_2^2 \, m_{++}^2
\ea

As we see, for all possible cases, when $N2$ is a minimum on always obtains $V_{CBi} - V_{N2} \,>\,0$.
This conclusion holds even if one considers complex charge breaking vevs, as is shown in Appendix~\ref{ap:nsbc},
and thus the stability of $N2$ against charge breaking seems to be guaranteed
 -- {\em provided that $c_1\neq 0$, our underlying (and subtle) assumption; if this
condition is relaxed, as will be shown in section~\ref{sec:sher}, this will change}.

\subsection{Stability of minima of type $N1$ against charge breaking}

The analysis of the previous section can be adapted trivially to $N1$ stationary points, and once again
one obtains analytical expressions relating the depth of the potential at $N1$ and CB extrema:
\ba
V_{CB1} - V_{N1} &=& \dfrac{c_{3}^{2} \, m_{+}^{2}}{ 4 \left( 1+\dfrac{2 v_\Delta^2}{v_\Phi^2} \right) } \nonumber \\
 & & \nonumber \\
V_{CB2} - V_{N1} &=& \frac{1}{4} \, c_3^2 \, m_{++}^2 \nonumber \\
& & \nonumber \\
V_{CB3} - V_{N1} &=& \frac{m_+^2\, c_3^2}{4\left(1\,+\,\displaystyle{\frac{2\,v^2_\Delta}{v^2_\Phi}}\right)} \,+\, \frac{1}{4}\,c_4^2\,m_{++}^2 - \dfrac{1}{8}\lambda_{3}v_{\Delta}^{2}\dfrac{c_{3}^{2}c_{4}}{c_{2}}\nonumber \\
& & \nonumber \\
V_{CB4} - V_{N1} &=& \dfrac{c_1^2 \, m^{2}_{+}}{4 \left( 2+\dfrac{v_\Phi^2}{v_\Delta^2}\right) } \,+\, \dfrac{1}{8} \, c_2^2 \, m_{++}^{2} \nonumber \\
& & \nonumber \\
V_{CB5} - V_{N1} &=& \dfrac{c_1^2 \, m^{2}_{+}}{ 4 \left( 2+\dfrac{v_\Phi^2}{v_\Delta^2}\right) } \,+\, \dfrac{1}{8} \, c_2^2 \, m_{++}^{2} \,+\, \dfrac{c_{3}^{2} \, m_{+}^{2}}{2 \left( 1+\dfrac{2 v_\Delta^2}{v_\Phi^2}\right) } \nonumber \\
& & \nonumber \\
V_{CB6} - V_{N1} &=& \dfrac{c_1^2 \, m^{2}_{+}}{ 2 \left( 2+\dfrac{v_\Phi^2}{v_\Delta^2}\right) } \,+\, \dfrac{c_{2}^{2} \, m_{+}^{2}}{2 \left( 1+\dfrac{2 v_\Delta^2}{v_\Phi^2}\right) } \,,
\label{eq:CBN1}
\ea
where the scalar masses appearing in these expressions are now evaluated at an $N1$ extremum, {\em i.e.} given by
eqs.~\eqref{eq:mch1} and~\eqref{eq:mchch1}. The conclusions we can draw here are the same for the $N2$ case:
if $N1$ is a minimum then all of its squared masses will be positive, and hence all the above potential differences
have positive values, except for the $CB3$ extrema. In this case, the simultaneous occurrence of both
$N1$ and $CB3$ extrema is only possible if $\lambda_{3}=\lambda_{5}=0$, which implies $m_{+}^2=m_{++}^2=0$. This
 leads to the degeneracy of both extrema, $V_{CB3} - V_{N1} = 0$. This of course means that such coexistence of
 extrema implies that for such parameter choices $CB3$ ceases to be charge-breaking -- in fact, without the $\lambda_3$ and
 $\lambda_5$ terms in the potential, it becomes possible to perform two independent $SU(2)$ transformations
 on the doublet and triplet, and thus ``rotate away" the charge breaking vevs of the triplet, transforming a
 seeming $CB3$ vacuum into the $N1$ one. The upshot, of course, is that for generic scalar potential parameters
 where neither $\lambda_3$ nor $\lambda_5$ are zero, there will be no $CB3$ extrema coexisting with $N1$.

Thus $N1$ minima are stable against the possibility of deeper charge breaking minima occurring. This conclusion,
like for the $N2$ case, also holds if one considers complex CB vevs. We present the respective expressions
for potential differences in Appendix~\ref{ap:nsbc} -- but again, for both real and complex vevs, we have assumed
that in the CB vacua the doublet has a vev, $c_1 \neq 0$. As we will see in section~\ref{sec:sher}, relaxing
that assumption will significantly change these conclusions.

\subsection{Stability of minima of type $N2$ against neutral extrema}

We have seen that $N2$, the phenomenologically-appealing minimum of the Higgs-triplet model
without the soft breaking term, is entirely stable against the possibility of deeper
charge breaking extrema. But of course, there are three types of neutral minima allowed
in this model, and therefore we might also have the possibility of an $N2$ minimum
having deeper $N1$ or $N3$ extrema. To clarify notation, for the purpose of this
section only let us call $\{v_1,v_2\}$ the doublet and triplet neutral vevs of the case
$N1$, and $v_3$ the triplet vev for the $N3$ case. The same bilinear formalism of previous sections
can be used to study the interplay between $N$ extrema, and following the procedure outlined
above we easily obtain
\be
V_{N1} - V_{N2} \,=\, \frac{1}{4} \,v_2^2\,m^2_{H,A}  \,.
\ee
Thus, the existence of an $N2$ minimum also implies that no deeper $N1$ extremum can occur. The situation
changes when one considers coexisting $N2$ and $N3$ extrema. For $N3$ we will have
\be
X_{N3} \,=\, \frac{1}{2}\,\left(\begin{array}{c} 0 \\ v^2_3 \\ 0 \\ 0 \\ 0
  \end{array}\right)\quad\quad ,\quad\quad
V^\prime_{N3}\,=\,A\,+\,B\,X_{N3}\,=\,
\left(\begin{array}{c} m^2_{H3} \\ 0 \\ \times \\
\times \\ 0
  \end{array}\right),
  \label{eq:xn3vln3}
\ee
where by ``$\times$" we represent a non-zero entry which will not be relevant for the calculation we
are undertaking here, and $m^2_{H3} = m^2 + (\lambda_2 + \lambda_3) \,v^2_3/2$ is the squared
scalar mass arising, in the $N3$ extremum, from the neutral doublet components~\footnote{Notice that
for $N3$ the pattern of symmetry breaking is very different from previous extrema; for instance, the
Goldstone bosons arise from triple field components, not doublet ones.}. Using eqs.~\eqref{eq:xn3vln3}
and~\eqref{eq:xn2vln2} and following the now usual procedure of performing alternate products between
vectors $X$ and $V^\prime$, it is easy to obtain the following relationship between the depths of
the potential at $N2$ and $N3$:
\be
V_{N3} - V_{N2} \,=\, \frac{1}{4} \,\left[v^2_\phi\,m^2_{H3} \,-\, v_3^2 \,m^2_{H,A} \right]
\,=\,
\frac{1}{4} \,\left( \frac{M^4}{\lambda_2 + \lambda_3}\,-\,\frac{m^4}{\lambda_1}\right)
 \,.
 \label{eq:N3N2}
\ee
It is then clear that $N2$ being a minimum does {\em not} guarantee it is the deepest one.
Depending on the values of the parameters of the potential, it may well happen that
$N2$ is a local minimum, with a deeper $N3$ extremum.

\section{Potential with soft-breaking term}
\label{sec:sb}

The introduction of the soft breaking term $\mu$ changes many things. Phenomenologically, the $N2$
vacuum ceases to be possible -- the minimisation conditions have no solution with $v_\Delta = 0$ when
$v_\phi \neq 0$. As we see from eq.~\eqref{eq:mA1}, the $N1$ vacuum no longer implies a massless
scalar, rather that state has a mass directly proportional to $\mu$. And as we will now show, the
soft breaking term has a significant impact in the stability of neutral vacua. Let us begin by
recalling that the $\mu$ term in the potential is cubic in the fields. The potential is therefore
no longer a sum of quadratic and quartic terms, it has a cubic contribution, $V_3$. Then,
eq.~\eqref{eq:homog} must be generalised and becomes
\be
\mbox{At any stationary point:}\, \frac{\partial V}{\partial \varphi_i}\,=\,0
\,\Longrightarrow\,\sum_i \varphi_i\,\frac{\partial V}{\partial \varphi_i}\,=\,0\,\Longrightarrow
2\,V_2\,+\,3\,V3\,+\,4\,V_4\,=\,0\,,
\ee
and hence the value of the potential at a given stationary point will now be given by,
instead of eq.~\eqref{eq:VSP},
\be
V_{SP}\;=\;\frac{1}{2}\,\left(V_2\right)_{SP}\,+\,\frac{1}{4}\,\left(V_3\right)_{SP}\;=\;
-\,\frac{1}{2}\,\left(V_3\right)_{SP}-\,\left(V_4\right)_{SP}\,.
\label{eq:VSP3}
\ee
The cubic contribution to the value of the potential isn't particularly complicated; for $N1$
and $CB1$ vevs, for instance, it is given by
\ba
\left(V_3\right)_{N1} &=& -\,\frac{\mu}{\sqrt{2}}\,v_\phi^2\,v_\Delta\, \nonumber\\
\left(V_3\right)_{CB1} &=& -\,\frac{\mu}{\sqrt{2}}\,c_1^2\,c_2\, .
\label{eq:V3}
\ea
However, such cubic terms in the vevs mean that the application of the bilinear formalism is not
at all obvious, since it relies of quadratic-plus-quartic potentials being easily expressed
as polynomials of quadratic field/vev variables. Nevertheless, with some ingenuity, we can
follow the steps outlined for the non soft breaking cases and adapt the demonstrations to include
the cubic terms when necessary. As before, we will explicitly perform one of the calculations
relating the value of the potential at two pairs of extrema and leave the remaining
demonstrations as an exercise to the reader, all the while showing the final results.

Our starting point is, as always, the hypothesis that the potential's minimisation conditions
allow coexisting solutions, in this case of types $N1$ and $CB1$.
Let us keep the definitions of the vectors $A$, $X$ and $V^\prime$ and the matrix $B$ from
eqs.~\eqref{eq:def},~\eqref{eq:Vl} and~\eqref{eq:xn1}. In particular, $X_{N1}$ will still
be given as in eq.~\eqref{eq:xn1} and we will have
\be
V^\prime_{N1}\,=\,A\,+\,B\,X_{N1}\,=\,
\left(\begin{array}{c} M^2_\Delta\,\frac{2 v^2_\Delta}{v_\Phi^2} \\ M^2_\Delta\, \\ -\frac{1}{4}\lambda_5\,v_\phi^2 \\
-\frac{1}{2}\lambda_5\,v_\phi^2 -\,\lambda_3 v_\Delta^2 \\ 0
  \end{array}\right)\,.
\ee
where $M^2_\Delta$ is defined in eq.~\eqref{eq:MD}. For a $CB1$ extremum, with vevs such as those of
eq.~\eqref{eq:CB1}, the vector $X_{CB1}$ is still given by the same expression as in eq.~\eqref{eq:xcb1vlcb1},
but $V^\prime_{CB1}$ is now greatly changed:
\be
V^\prime_{CB1}\,=\,A\,+\,B\,X_{CB1}\,= \left(\begin{array}{c} M^2_\Delta\,\frac{2 c_2\, v_\Delta }{v_\Phi^2} \\
M^2_\Delta\,\frac{c_1^2 \,v_\Delta }{c_2\,v_\Phi^2}\\ 0 \\ -\frac{1}{2}\lambda_5 c_1^2
- \lambda_3 c_2^2 \\ -\lambda_3 c_3^2
  \end{array}\right)\,,
\ee
where for convenience we are using the $N1$-related $M_\Delta$ and vevs. We now perform the product between
vectors $X_{CB1}$ and $V^\prime_{N1}$, obtaining
\be
X_{CB1}^T\,V^\prime_{N1}\,=\,A^T\,X_{CB1}\,+\,X_{CB1}^T\,B\,X_{N1}\,=\,
\frac{1}{2}\,\left[ M^2_\Delta \left( 2\frac{v_\Delta^2}{v_\Phi^2} c_1^2\,+\,c_2^2 \right)\,+\,
\left( M^2_\Delta  \,-\, \frac{1}{4} \,\lambda_5 \,v_\Phi^2 \right)
\,c_3^2 \right]\,.
\ee
The term multiplying $c_3^2$ in this expression is directly proportional to the squared charged mass at the
$N1$ extremum, {\em viz.} eq.~\eqref{eq:mch1}. As for the term $A^T\,X_{CB1}$ it gives us the value of
the quadratic terms of the potential at the $N1$ stationary point, so that, using eqs.~\eqref{eq:VSP3}
and~\eqref{eq:V3}, we can rewrite it as
\be
A^T\,X_{CB1}\,=\,2 \,V_{CB1}\,+\,\frac{\mu}{2\sqrt{2}}\,c_1^2\,c_2 \,=\,2 \,V_{CB1}\,+\,M^2_\Delta
\frac{v_\Delta}{2v_\Phi^2}\,c_1^2\,c_2\,.
\ee
We therefore see the value of the potential at the $CB1$ extremum appearing, and we rewrote the value of $\mu$
using eq.~\eqref{eq:MD} for later convenience. Thus we end up obtaining
\be
V_{CB1}\,=\,-\,\frac{1}{2}\,M^2_\Delta \frac{v_\Delta}{2v_\Phi^2}\,c_1^2\,c_2
-\frac{1}{2}\,X_{CB1}^T\,B\,X_{N1} +
\frac{1}{4}\,\left[M^2_\Delta \left(2\frac{v_\Delta^2}{v_\Phi^2} c_1^2+c_2^2\right)+
\left(M^2_\Delta  -\frac{1}{4} \,\lambda_5 \,v_\Phi^2 \right)\,c_3^2 \right]\,.
\label{eq:Vcb1nsb}
\ee
Likewise, the product $X_{N1}$ and $V^\prime_{CB1}$ gives us
\be
X_{N1}^T\,V^\prime_{CB1}\,=\,A^T\,X_{N1}\,+\,X_{N1}^T\,B\,X_{CB1}\,=\,
\frac{1}{2}\, M^2_\Delta \left( 2 v_\Delta c_2 \,+\, \frac{c_1^2 v_\Delta^3}{v_\Phi^2 c_2} \right)\,,
\ee
and from eqs.~\eqref{eq:VSP3} and~\eqref{eq:V3} we also obtain
\be
A^T\,X_{N1}\,=\,2 \,V_{N1}\,+\,\frac{\mu}{2\sqrt{2}}\,v_\Phi^2\,v_\Delta \,=\,2 \,V_{N1}\,+\,
\frac{1}{2}\,M^2_\Delta v_\Delta^2\,.
\ee
And thus,
\be
V_{N1}\,=\,-\,\frac{1}{2}\,M^2_\Delta v_\Delta^2-\frac{1}{2}\,X_{N1}^T\,B\,X_{CB1} \,+\,
\frac{1}{2}\, M^2_\Delta \left( 2 v_\Delta c_2 \,+\, \frac{c_1^2 v_\Delta^3}{v_\Phi^2 c_2} \right)\,.
\label{eq:Vn1nsb}
\ee
We can now subtract eqs.~\eqref{eq:Vcb1nsb} and~\eqref{eq:Vn1nsb} and, using
the fact that the matrix $B$ is symmetric and eq.~\eqref{eq:mA1}, after some algebra we finally obtain
\be
V_{CB1} - V_{N1} \,=\, \frac{m^2_A}{4\left(1\,+\,\displaystyle{\frac{4\,v^2_\Delta}{v^2_\Phi}}\right)}\,
(c_2 - v_\Delta)^2\,\left( 1\,-\,\frac{v_\Delta}{c_2}\,\frac{c_1^2}{v^2_\Phi} \right) \,+\,
\frac{m_+^2\, c_3^2}{4\left(1\,+\,\displaystyle{\frac{2\,v^2_\Delta}{v^2_\Phi}}\right)}\, .
\ee
A few comments are in order while analysing this expression:
\begin{itemize}
\item As before, the difference in the values of the potentials at a CB stationary point and
a $N1$ one can be expressed as a function of vevs and the squared masses at $N1$.
\item If one takes the limit $\mu\rightarrow 0$ in this expression (equivalent to making $m_A = 0$)
one recovers the non-soft breaking expression of eq.~\eqref{eq:CBN1}.
\item Unlike the $\mu = 0$ case, however, now even if $N1$ is a minimum, rendering both $m^2_A$
and $m_+^2$ positive, it is no longer guaranteed that $V_{CB1} - V_{N1} > 0$.
\item The reason is the minus sign affecting the coefficient $v_\Delta/c_2$, which opens up the
possibility of having $V_{CB1} - V_{N1} < 0$ even if $N1$ is a minimum.
\end{itemize}

Thus the soft breaking coefficient $\mu$ completely changes the stability properties of the $N1$ minimum.
In fact, performing similar calculations to those detailed for the $CB1$ case for the remaining $CB$
allowed vacua (with real vevs), it is possible to find that, for $N1$ coexisting with other $CB$
extrema, one has
\ba
V_{CB2} - V_{N1} &=& \frac{m^2_A}{4\left(1\,+\,\displaystyle{\frac{4\,v^2_\Delta}{v^2_\Phi}}\right)}\,(c_2 - v_\Delta)^2\,\left( 1\,-\,\frac{v_\Delta}{c_2}\,\frac{c_1^2}{v^2_\Phi} \right) \,+\, \frac{1}{4}\,c_3^2\,m_{++}^2\nonumber \\
 & & \nonumber \\
V_{CB3} - V_{N1} &=& \frac{m^2_A}{4\left(1\,+\,\displaystyle{\frac{4\,v^2_\Delta}{v^2_\Phi}}\right)}\,(c_2 - v_\Delta)^2\,\left( 1\,-\,\frac{v_\Delta}{c_2}\,\frac{c_1^2}{v^2_\Phi} \right) +\,\frac{m_+^2\, c_3^2}{4\left(1\,+\,\displaystyle{\frac{2\,v^2_\Delta}{v^2_\Phi}}\right)} \,+\, \frac{1}{4}\,c_4^2\,m_{++}^2 - \dfrac{1}{8}\lambda_{3}v_{\Delta}^{2}\dfrac{c_{3}^{2}c_{4}}{c_{2}}\nonumber \\
 & & \nonumber \\
V_{CB4} - V_{N1} &=& \frac{m^2_A}{4\left(1\,+\,\displaystyle{\frac{4\,v^2_\Delta}{v^2_\Phi}}\right)}\,\left( \dfrac{c_{2}^{2}}{2} + v_\Delta^2 + c_{1}^{2}\dfrac{v_{\Delta}^2}{v_{\Phi}^2}\right) \, + \, \dfrac{1}{8} c_{2}^{2} m_{++}^{2} \, + \,\dfrac{v_{\Delta}^2}{v_{\Phi}^2}\dfrac{c_{1}^2 \, m_{+}^2}{4\left(1\,+\,\displaystyle{\frac{2\,v^2_\Delta}{v^2_\Phi}}\right)}\nonumber \\
 & & \nonumber \\
V_{CB5} - V_{N1} &=& \frac{m^2_A}{4\left(1\,+\,\displaystyle{\frac{4\,v^2_\Delta}{v^2_\Phi}}\right)}\,\left( \dfrac{c_{2}^{2}}{2} + v_\Delta^2 + c_{1}^{2}\dfrac{v_{\Delta}^2}{v_{\Phi}^2}-c_{3}^2\right) \, + \, \dfrac{1}{8} c_{2}^{2} m_{++}^{2} \, + \,\dfrac{m_{+}^2}{4\left(1\,+\,\displaystyle{\frac{2\,v^2_\Delta}{v^2_\Phi}} \right)} \left( c_{1}^{2}\dfrac{v_{\Delta}^2}{v_{\Phi}^2} + 2\,c_{3}^2 \right)\nonumber \\
& & \nonumber \\
V_{CB6} - V_{N1} &=& \frac{m^2_A}{4\left(1\,+\,\displaystyle{\frac{4\,v^2_\Delta}{v^2_\Phi}}\right)}\,\left( v_\Delta^2 - c_{2}^2\right) \, + \,\dfrac{m_{+}^2}{2\left(1\,+\,\displaystyle{\frac{2\,v^2_\Delta}{v^2_\Phi}} \right)} \left( c_{1}^{2}\dfrac{v_{\Delta}^2}{v_{\Phi}^2} + c_{2}^2 \right)\, .
\ea
Only the $CB4$ case is guaranteed to give $V_{CB4} - V_{N1} > 0$ when $N1$ is a minimum. In all other five cases,
there is always at least one negative term somewhere in the expressions that can render the potential differences negative
even when $N1$ is a minimum. The inclusion of complex vevs only reinforces this conclusion (see
Appendix~\ref{ap:sbc}). Thus, ``turning on" the soft breaking term in the potential weakens the stability of
neutral minima -- even if $N1$ is a minimum, there may be regions of parameter space for which deeper charge breaking
vacua with $c_1 \neq 0$ occur.

\section{The case of the vevless doublet}
\label{sec:sher}

Up until this point we have been considering only CB vev configurations with $c_1 \neq 0$, that is, the
doublet always possessing a vev. Consider, however, that the first derivative of the potential with
respect to $c_1$, from eq.~\eqref{eq:VCB3}, is given by
\be
\frac{\partial V}{\partial c_1}\,=\,c_1\,\left[
m^2 \,+\,\lambda_1 \,c_1^2 \,+\,
\frac{\lambda_4}{2} (c_2^2 +  c_3^2 +  c_4^2) \,+\,
\frac{\lambda_5}{2} \, (2 c_2^2 +  c_3^2) \right]\,=\,0\, .
\ee
From here we see that the trivial solution $c_1 = 0$ is {\em always} possible, regardless of the
values of the parameters. Not only that, it is a disconnected solution from $c_1\neq 0$ -- the latter
solution imposes a relation between CB vevs and scalar potential parameters, whereas the former one
does not. This means that the conclusions we drew for CB vacua with $c_1\neq 0$ cannot be extended
to the vevless doublet case by taking the limit $c_1 \rightarrow 0$. Thus there is the possibility
that the $c_1 = 0$ case brings qualitatively different conclusions, and indeed that will be the case, as we will
now show.

With $c_1=0$, there are six new possible real CB vev choices, which we will dub cases $CB7$ to $CB12$,
for which the minimisation equations give non-trivial solutions:
\ba
\langle \Phi\rangle_{CB7} = \frac{1}{\sqrt{2}}\,\left(
\begin{array}{c} 0 \\ 0 \end{array}
\right)\;\;\; &,& \;\;\; \langle \Delta \rangle_{CB7} = \frac{1}{\sqrt{2}}\,
\left( \begin{array}{cc} c_3/\sqrt{2} & c_2 \\
c_2 & -c_3/\sqrt{2} \end{array}  \right)
\label{eq:CB7} \\
\langle \Phi\rangle_{CB8} = \frac{1}{\sqrt{2}}\,\left(
\begin{array}{c} 0 \\ 0 \end{array}
\right)\;\;\; &,& \;\;\; \langle \Delta \rangle_{CB8} = \frac{1}{\sqrt{2}}\,
\left( \begin{array}{cc} 0 & c_2 \\
c_2 & 0 \end{array}  \right)
\label{eq:CB8} \\
\langle \Phi\rangle_{CB9} = \frac{1}{\sqrt{2}}\,\left(
\begin{array}{c} 0 \\ 0 \end{array}
\right)\;\;\; &,& \;\;\; \langle \Delta \rangle_{CB9} = \frac{1}{\sqrt{2}}\,
\left( \begin{array}{cc} 0 & -c_2 \\
c_2 & 0 \end{array}  \right)
\label{eq:CB9} \\
\langle \Phi\rangle_{CB10} = \frac{1}{\sqrt{2}}\,\left(
\begin{array}{c} 0 \\ 0 \end{array}
\right)\;\;\; &,& \;\;\; \langle \Delta \rangle_{CB10} = \frac{1}{\sqrt{2}}\,
\left( \begin{array}{cc} c_3/\sqrt{2} & -c_3^2/2c_2 \\
c_2 & -c_3/\sqrt{2} \end{array}  \right)
\label{eq:CB10} \\
\langle \Phi\rangle_{CB11} = \frac{1}{\sqrt{2}}\,\left(
\begin{array}{c} 0 \\ 0 \end{array}
\right)\;\;\; &,& \;\;\; \langle \Delta \rangle_{CB11} = \frac{1}{\sqrt{2}}\,
\left( \begin{array}{cc} 0 & c_4 \\
0 & 0 \end{array}  \right)
\label{eq:CB11} \\
\langle \Phi\rangle_{CB12} = \frac{1}{\sqrt{2}}\,\left(
\begin{array}{c} 0 \\ 0 \end{array}
\right)\;\;\; &,& \;\;\; \langle \Delta \rangle_{CB12} = \frac{1}{\sqrt{2}}\,
\left( \begin{array}{cc} c_3/\sqrt{2} & 0 \\
0 & -c_3/\sqrt{2} \end{array}  \right)
\label{eq:CB12}
\ea
We can now apply the same methodology of previous sections to the comparison of
the value of the potential at each of the above CB vacua and the normal ones.

\subsection{Stability of minima of type $N1$ and $N2$ against charge breaking without soft-breaking}

With $\mu = 0$ the potential has an intact global symmetry, as we discussed in section~\ref{sec:mod}. Previously
we concluded that without the soft breaking term there was no possibility of deeper CB vacua
{\em with $c_1 \neq 0$} than neutral ones. Now, however, the conclusions will differ:
\ba
V_{CB7} - V_{N1} &=& \frac{v^2_\Phi}{v_\Delta^2}\,\dfrac{m^2_h\,m^2_H}{16(\lambda_2+\lambda_3)} \, - \, \lambda_{3}\dfrac{[2(\lambda_2+\lambda_3)v_\Delta^2+(\lambda_4+\lambda_5)v_\Phi^2]^2}{16(\lambda_2+\lambda_3)(2\lambda_2+\lambda_3)}\nonumber \label{eq:VCB7} \\
 & & \nonumber \\
V_{CB10} - V_{N1} &=& \frac{v^2_\Phi}{v_\Delta^2}\,\dfrac{m^2_h\,m^2_H}{16(\lambda_2+\lambda_3)}\, .
\label{eq:VCB10}
\ea
The expressions for $V_{CB7} \, - \, V_{N1}$ holds for $CB8$, $CB9$ and $CB12$, while the second one also holds for $V_{CB11} - V_{N1}$. We see from eq.~\eqref{eq:VCB10} that an $N1$ minimum is stable against deeper vacua
$CB10$ (and $CB11$), due to the boundedness-from-below conditions of eq.~\eqref{eq:bfb} ensuring that the quantity
$\lambda_2+\lambda_3$ in eq.~\eqref{eq:VCB10} is positive. However, eq.~\eqref{eq:VCB7} tells us that, even if $N1$ is
a minimum, there is no guarantee that $V_{CB7} - V_{N1}\,>\,0$ -- in fact, though the first term in the right-hand-side
of eq.~\eqref{eq:VCB7} is certainly positive if $N1$ is a minimum, the same cannot be said for the second term,
which can have either sign. Thus in fact the $N1$ minima can be unstable against CB when the soft breaking term is
absent, but only for CB vacua for which only the triplet has vevs. This completely changes the stability properties
of this version of the HTM -- if one analysed only CB vacua with a vev for the doublet, neutral minima in
this HTM with a global symmetry were seemingly CB-stable, but in fact deeper CB vacua with vevless
doublet are possible.

For $N2$ minima, we will have:
\ba
V_{CB7} - V_{N2} &=&  \frac{1}{4} \,\left(\frac{m^4}{\lambda_1} \,-\,\frac{M^4}{\lambda_2 +
\dfrac{1}{2}\lambda_3}\right) \nonumber \\
 & & \nonumber \\
 V_{CB10} - V_{N2} &=&  \frac{1}{4} \,\left(\frac{m^4}{\lambda_1} \,-\,\frac{M^4}{\lambda_2 +
\lambda_3}\right)
\label{eq:N2CB0}
\ea
where we recognise formulae similar to eq.~\eqref{eq:N3N2}. More importantly, these
expressions confirm that minima of type $N2$ are {\em not} guaranteed to be stable against charge
breaking -- depending on the values of the parameters of the potential, deeper CB vacua with
$c_1 = 0$ may well exist.

\subsection{Stability of minima of type $N1$ against charge breaking with soft-breaking}

Finally, considering now the case of the potential with a soft breaking term $\mu$,
CB vacua with a vevless doublet can also occur, and its relationship with the vacuum
$N1$ are such that:
\ba
V_{CB7} - V_{N1} &=& \frac{v^2_\Phi}{v_\Delta^2}\,\dfrac{m^2_h\,m^2_H}{16(\lambda_2+\lambda_3)} \, - \dfrac{\lambda_1}{8(\lambda_{2}+\lambda_3)}\dfrac{m_A^2}{1+\dfrac{4v_\Delta^2}{v_\Phi^2}}\dfrac{v_\Phi^4}{v_\Delta^2} \, - \, \lambda_{3}\dfrac{[2(\lambda_2+\lambda_3)v_\Delta^2+(\lambda_4+\lambda_5)v_\Phi^2]^2}{16(\lambda_2+\lambda_3)(2\lambda_2+\lambda_3)} \nonumber \\
& & \nonumber \\
&+& \, \frac{\lambda_{3}}{2(2\lambda_2+\lambda_3)} \, \dfrac{m_A^2}{1+\dfrac{4v_\Delta^2}{v_\Phi^2}} \left[
v_\Delta^2 + \dfrac{\lambda_4 + \lambda_5}{2(\lambda_2+\lambda_3)} v_\Phi^2 -
\dfrac{1}{2(\lambda_2+\lambda_3)} \dfrac{m_A^2}{1+\dfrac{4v_\Delta^2}{v_\Phi^2}} \right]  \nonumber \\
 & & \nonumber \\
V_{CB10} - V_{N1} &=& \frac{v^2_\Phi}{v_\Delta^2}\,\dfrac{m^2_h\,m^2_H}{16(\lambda_2+\lambda_3)} \, - \, \dfrac{\lambda_1}{8(\lambda_{2}+\lambda_3)} \, \dfrac{m_A^2}{1+\dfrac{4v_\Delta^2}{v_\Phi^2}} \, \dfrac{v_\Phi^4}{v_\Delta^2}\, .
\ea
The expression for $V_{CB7} \, - \, V_{N1}$ holds for the cases $CB8$, $CB9$ and $CB12$, while the second one also holds for $V_{CB11} - V_{N1}$. Again we conclude that the fact that $N1$ is a minimum does not guarantee its stability
against deeper charge breaking vacua. However, in this softly-broken model we had already identified, in
section~\ref{sec:sb}, CB vacuum configurations for which deeper CB minima could coexist with neutral ones.
Thus for the softly broken potential the vevless doublet case does not bring any qualitatively different conclusions.

\section{Numerical analysis}
\label{sec:num}

To ascertain the relevance of the previous results, we will now undertake a numerical analysis of the
parameter space of the Higgs Triplet Model, searching for CB minima deeper than neutral ones. The aim
is to verify whether restrictions on the model's parameters can be obtained by requiring that the global
minimum of the model be neutral, thus increasing its predictive power.

\begin{figure}
  \centering
  \includegraphics[height=8cm,angle=0]{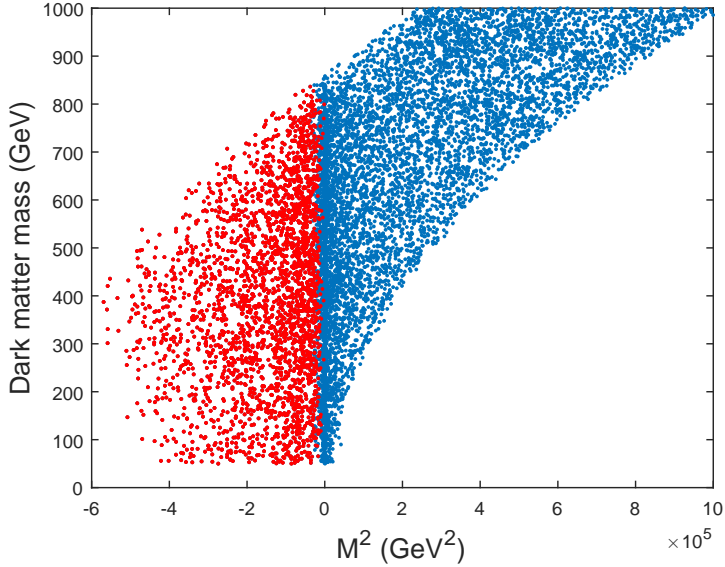}
  \caption{Values of the dark matter particle mass as a function of the quadratic coupling $M^2$ for
  a minimum of type $N2$. In blue, all the scanned points; in red, those points for which there exists
  a CB vacuum (of types $CB7$ or $CB10$) lower than $N2$.}
  \label{fig:N2}
\end{figure}

We begin with the HTM with the global symmetry intact, without the soft breaking term $\mu$. As discussed
earlier, this model has a vaccum of type $N2$, given by eq.~\eqref{eq:N2}, which includes possible dark
matter candidates -- the CP-even scalar $H$ or the pseudoscalar $A$, degenerate in mass. We generated a large sample
(10000 points) of combinations of parameters satisfying the following conditions:
\begin{itemize}
\item The SM-like Higgs boson has a mass of 125 GeV; the remaining scalar masses were chosen randomly
in the intervals
\be
50 \leq m_H\,=\,m_A\,\leq 1000\,\mbox{GeV}\;\;, \;\; \mbox{max}\{m_H\,,\,400\} \leq m_+ \,,\,m_{++}
\leq 1000\,\mbox{GeV}\,.
\ee
\item The quartic couplings $\lambda_2 $ and $\lambda_3$ are chosen randomly and independently in the
interval $[-10\,,\,10]$.
\item The quadratic parameter $M^2$ is chosen randomly in the interval $[-10^6 \,,\,10^5]$ GeV$^2$.
\end{itemize}
These choices do not pretend to be an exhaustive scan of the model's parameter space -- we merely wish
to show that,
for regions of parameter space which may be of phenomenological interest, bounds arising from requiring no
deeper CB minima are relevant. We chose the dark matter masses to include both the case where it is
reasonably light
(tens of GeV) or fairly heavy (up to 1 TeV), and required that it is the lightest of the scalars stemming
from the triplet. The equations~\eqref{eq:mh2}--\eqref{eq:mchch2} relate the masses and the couplings at this
$N2$ minimum, and allow us to fully specify all parameters of the potential. We then required that the quartic
couplings obeyed the bounded from below and unitarity conditions described in
section~\ref{sec:mod}. Our choice
for the masses of the charged particles is a simple way to ensure that the scalar contributions
to the diphoton decay
of $h$ are not too large, and therefore $h$ behaves, in all of its production and decay channels, very much
like the SM Higgs boson, as current LHC results indicate is the case~\footnote{Notice that, due to the intact
global symmetry, the $h$ scalar has tree-level couplings to fermions and gauge bosons identical
to those of the
SM. We can therefore be confident that the chosen parameter space yields a 125 GeV scalar with properties
in numerical agreement with LHC results.}. Once the parameter space was generated we searched
for charge breaking
minima. That search could have been done by a numerical minimisation of the full HTM scalar
potential (which, remember,
depends on 10 real scalar component fields), but the practical and useful aspect of our work consisted in
identifying the most likely CB vacua -- in this case, the vev combinations
we dubbed $CB7$ and $CB10$, in eqs.~\eqref{eq:CB7}, \eqref{eq:CB10}. Both of them yield
relatively straightforward
equations which permit to determine the values of the CB vevs. Once we have chosen all parameters of the
potential which yield an $N2$ minimum, the relations of those parameters with the CB vevs are:
\ba
CB7:\; & &  c_3^2 + 2 c_4^2 \,=\,-\,\frac{M^2}{\lambda_2 + \dfrac{1}{2}\lambda_3}\;\;\; , \;\;\; c_2\,=\,c_4
\nonumber\\
CB10:\; & &  \frac{\left(2 c_2^2 + c_3^2\right)^2 }{4 c_2^2} \,=\,-\,\frac{M^2}{\lambda_2 + \lambda_3}\;\;\; ,
\;\;\; c_4\,=\,-\,\frac{c_3^2}{2 c_2}\,.
\label{eq:cbvs}
\ea
The procedure we followed was therefore quite simple: for each set of parameters $M^2$,
$\lambda_2$, $\lambda_3$,
choose random values for the CB vevs $c_i$ such that the equations above were satisfied; with those values of
the CB vevs it was then a simple matter to compute the value of the potential at the CB extremum
and compare it
with its value at the $N2$ minimum. We present the results of this procedure in fig.~\ref{fig:N2} --
in the plot we see the distribution of the parameter points in the $M^2$-$m_H$ plane (recall that for this
minimum $H$ and $A$ have degenerate masses and are dark matter candidates). The blue points are the totality
of the scan -- the red points are a subset of the blue ones, and indicate the regions of parameter space
for which there is a $CB$ vacuum (of types $CB7$ or $CB10$) lower than the $N2$ minimum. We see some interesting
features emerging from this plot:
\begin{itemize}
\item The deeper $CB$ vacua can only occur if $M^2 < 0$. This is easily understood from
eqs.~\eqref{eq:cbvs},
and if we recall that, since the bounded from below conditions of eqs.~\eqref{eq:bfb} are satisfied,
the quantities $\lambda_2 + \lambda_3$ and $\lambda_2 + \lambda_3/2$ are positive. Existence of a $CB$
extremum therefore requires negative $M^2$.
\item For all points with positive $M^2$ the $N2$ minimum is global. Nonetheless, we also
 observe that $M^2>0$
is a {\em sufficient} condition for $N2$ stability, not a {\em necessary} one -- there are certainly blue
points in the region $M^2 < 0$.
\item Likewise, points with a very high dark matter mass (above roughly 840 GeV) are safe from $CB$
instability. But once more, requiring $m_H\,>\, 840 $ GeV would be a {\em sufficient} condition to ensure
the non-existence of deeper $CB$ vacuua, not a {\em necessary} one.
\item That $CB$ vacua occurs for lower dark matter masses of $H$ or $A$ is simple to understand
if one considers
eq.~\eqref{eq:mH2}: there we see that the
\be
m^2_H = m^2_A \, = \, M^2 + \frac{1}{2} (\lambda_4 + \lambda_5) v^2
\ee
and since $CB$ vacua need $M^2 < 0$ and the magnitude of the $\lambda$ couplings is limited by
the unitarity of the theory, if follows naturally that in the region of parameter space where
deeper $CB$ vacua might occur the dark matter masses will tend to be smaller.
\end{itemize}
In all, roughly 27\% of the scanned parameter space includes global $CB$ vacua. Of course, that they
are deeper does not mean that these $CB$ vacua are necessarily dangerous -- one would need to compute
the tunneling time between $N2$ and the global minima to verify whether it is smaller than the age of
the universe. But the simple fact that such a large percentage of parameter points have a deeper $CB$ vacuum
is sobering. And analysing directly the expressions relating the relative depth of the potential
at $N2$ and $CB7$ or $CB10$ (eqs.~\eqref{eq:N2CB0}) we can easily deduce a necessary and sufficient
condition for non-existence of deeper CB vacua: from that equation we see that the condition to have
$V_{CB} - V_N >0$ is
\be
M^4\,<\,\mbox{min}\left(\lambda_2 + \dfrac{1}{2} \lambda_3\,,\,\lambda_2 + \lambda_3\right)\,\frac{m^4}{\lambda_1}\,.
\ee
But remember that eqs.~\eqref{eq:N2CB0} were deduced assuming the existence of a CB extremum, which requires
$M^2 < 0$ -- for all positive values of $M^2$, neutral vacua stability is guaranteed. Thus we can obtain
from the above equation that the neutral minimum $N2$ is stable against deeper charge breaking vacua
if and only if
\be
M^2\,>\,-\,\sqrt{\mbox{min}\left(\lambda_2 + \dfrac{1}{2} \lambda_3\,,\,\lambda_2 + \lambda_3\right)}
\,\frac{m^4}{\lambda_1}\,,
\ee
where we have used the positivity (imposed from bounded from below conditions) of the combinations of
$\lambda_2$ and $\lambda_3$ couplings under the square root. Further using the fact that at the $N2$
minimum one has $m^2 = -\lambda_1 v^2 = -m^2_h/2$, we obtain:
\be
\mbox{An $N2$ minimum is stable against charge breaking iff }\,
M^2\,>\,-\,\sqrt{\mbox{min}\left(\lambda_2 + \dfrac{1}{2} \lambda_3\,,\,\lambda_2 + \lambda_3\right)}
\,\frac{m_h\,v}{\sqrt{2}}\,.
\label{eq:bou}
\ee

Considering now a softly broken model with a minimum of type $N1$, the analysis of the previous sections
shows there are several types of possible deeper $CB$ vacua. We scanned over the model's parameter space,
allowing the triplet vev to be at most $\sim 8$ GeV, in order to comply with electroweak precision
constraints~\cite{Arhrib:2011uy,Arhrib:2011vc,Kanemura:2012rj,Aoki:2012yt,Aoki:2012jj}. We allowed
the quartic parameters $\{\lambda_2\,,\,\lambda_3\,,\,\lambda_4\}$ to vary between -10 and 10 and
used the expressions for the eigenvalues of the CP-even mass matrix to, through the input
of the values of $m_h$ and $m_H$, determine the quartic coupling $\lambda_5$ and the soft breaking
parameter $\mu$. With all the potential's parameters thus established we demanded that they obeyed
unitarity and boundedness from below conditions; and also that the phenomenology of the 125 GeV
scalar, $h$, be SM-like as per current LHC results -- to do this, we required that the tree-level couplings of
$h$ to gauge bosons and fermions (which can be found in table I in ref.~\cite{Aoki:2012jj}) be at most
10\% deviated from their expected SM values, which is a degree of precision even superior to the current one.
Once again, our purpose is not to perform a complete parameter space scan but rather show that CB bounds
are relevant to phenomenologically appealing regions of parameter space of the model.

\begin{figure}
  \centering
  \includegraphics[height=8cm,angle=0]{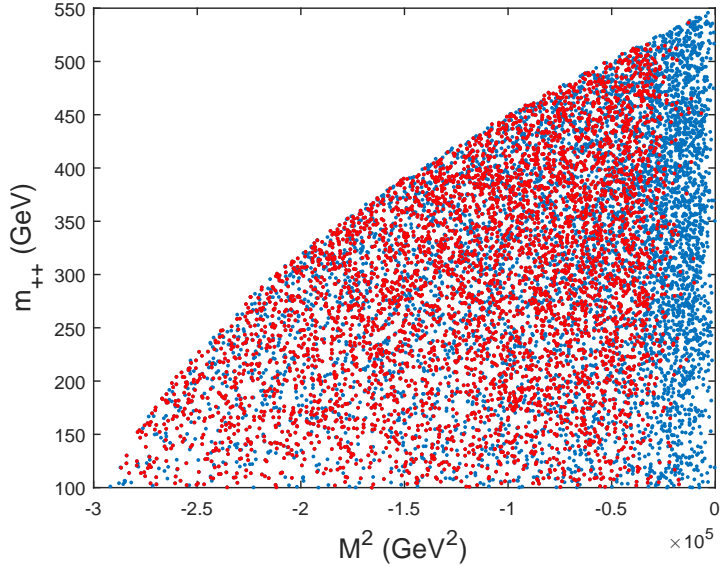}
  \caption{Values of the doubly charged scalar mass a function of the quadratic coupling $M^2$ for
  a minimum of type $N1$ in an HTM with softly broken global symmetry. In blue, all the scanned points; in red,
  those points for which there exists
  a CB vacuum (of types $CB7$ or $CB10$) lower than $N1$.}
  \label{fig:N1}
\end{figure}

The maximum values of the masses we find for the extra
scalars in this minimum are smaller than those we found for the $N2$ case -- this is a natural consequence
of the fact that in this case $M^2$ is directly related through the minimisation conditions of the potential
to the vevs $v_\Phi$ and $v_\Delta$, the soft breaking parameter $\mu$ and to the quartic couplings, whereas in $N2$
that parameter, which by and large determines the magnitude of the extra scalar masses, is not determined by the
minimisation of the potential. Having determined the full set of parameters caracterising an $N1$ minimum we
then proceed to verify whether there is a deeper $CB$ vacuum, by performing a numerical minimisation of the
potential whilst allowing the $CB$ vevs to be non-zero. The results of that procedure are shown in
fig.~\ref{fig:N1}, where we plot the doubly charged scalar mass $m_{++}$ (from eq.~\eqref{eq:mchch1})
as a function of $M^2$. As before, in blue we represent the entirety of scanned points, and in
red the subset -- a little over 48\% -- of those points for which there is a $CB$ vacuum below the
$N1$ minimum. As in the non-soft breaking $N2$ case we see that deeper $CB$ vacua occur exclusively
for $M^2<0$,
but now there is a substantial number of blue points in the $M^2<0$ region, ``in the middle" of the
red ones~\footnote{Obviously we are dealing with an 8 dimensional parameter space, of which we are
only showing a 2D slice. In other (hyper)planes the separation between $CB$ and neutral vacua might
be much sharper, but we found no such representations.}. In fact, the existence of the $\mu$ parameter
changes considerably the stability picture
of the neutral minima -- not only is there is a greater percentage of unstable $N1$ minima for the
potential which includes $\mu$, but also the
regions for which $N1$ stability is guaranteed are now quite different from the $N2$ case. But
such a large percentage
of potentially-unstable neutral minima shows that one needs to be careful when considering parameter
scans of the Higgs Triplet Model, lest the values of the parameters chosen actually predict a global
$CB$ minimum.

There is another potential instability for $N1$ minima -- the possibility that there exists a second
minimum of type $N1$ ($N1^\prime$), with different values for the vevs of the doublet and triplet. In fact,
the minimisation equations for an $N1$-type minimum admit, a priori, several solutions. We deal with
this intriguing possibility in appendix~\ref{ap:n1l}, but in practical terms it has no impact on our
results: for the whole of our scanned parameter space there is no $N1^\prime$ minimum such that
$V_{N1^\prime} < V_{N1}$.

\section{Conclusions}
\label{sec:con}

In this paper we analysed the stability of neutral minima in the Higgs Triplet Model against the
possibility of deeper charge breaking minima developing. We performed a thorough search of
possible CB vev configurations and found analytical expressions relating the difference in the
depths of the potential at neutral and CB extrema. We separated the analysis of two versions of the
model -- the model possessing a global symmetry and the model where that same symmetry is softly broken
by a cubic term. We also performed a separate study of the CB vev configurations with or without
a vev for the doublet. The analytical calculations helped us establish that, in some cases,
only the vevless doublet vacua could yield CB global minima. They also established that the
inclusion of the soft breaking term induces more possibilities of CB vacua developing,
changing the picture of stability of the model. It is to
be expected that the introduction of the soft-breaking coefficient $\mu$ changes the stability of the
potential -- the term with $\mu$ is a cubic one, and cubic terms in scalar potentials typically induce
vacuum instabilities (see, for instance, the SUSY
case~\cite{Frere:1983ag,Casas:1995pd,Ferreira:2000hg,Ferreira:2001tk}). To verify the relevance of
CB bounds one might obtain we performed a numerical scan over the parameter space of the model. We found that
for roughly 26\% (48\%) of the parameter space found for the globally symmetric (softly broken) potential
neutral minima had deeper charge breaking ones. For the dark matter minimum, there was a clear demarcation for the
 regions where CB could occur, not so for the softly broken model.
The potential for instability is therefore quite present, and
in principle tunneling calculations to the deeper vacua would become necessary -- though an alternative is to
simply exclude the combinations of parameter which produce a deeper CB vacuum, on the argument that thermal
fluctuations in the early universe increase immensely the probability that the model occupies the global
minimum, instead of becoming trapped in a local one~\cite{Barroso:2012mj,Barroso:2013awa,Hollik:2018wrr}.

The first general conclusion to draw from this work is that CB global minima can indeed coexist, in some
cases fairly frequently, with neutral minima. The authors of refs.~\cite{Arhrib:2011uy,Arhrib:2011vc}
identified regions of parameter space for which CB extrema were deeper than neutral ones and took them into
account in their phenomenological analysis of the model. Theirs was a partial and qualitative analysis,
but which already showed the likely importance that CB bounds could have. The remarkable work of
ref.~\cite{Xu:2016klg} showed that it was possible to obtain analytical expressions relating the
depths of the potential at different extrema. The expressions obtained therein were exact for the
potential with an intact global symmetry -- and we reproduce the results of that work for our vacua
comparisons -- and approximate when the soft breaking parameter $\mu$ was different
from zero. The author of~\cite{Xu:2016klg} was concerned with the use of the HTM to generate neutrino masses
via a Type-II Seesaw mechanism, thus obtaining approximate expressions for the relative potential depth in the
limit $\mu \rightarrow 0$ is certainly worthwhile. The expressions we present in the current work are
valid for any value of $\mu$. We also privilege writing the relations between potential depths, when possible,
in terms of scalar squared masses, thus automatically enlightening, in many cases, whether certain minima are global
or not. We also considered in greater detail the vevless doublet vacua, which we found to be relevant in many cases,
where indeed they were, for the $\mu = 0$ case, the only possible sources of vacuum instability for certain minima.

The current work organises the possibilities
of CB vacua which may be dangerous and may well simplify future numerical studies -- instead of blindly minimising
a 10-field potential, researchers can now look for specific (and thus dependent on less variables)
combinations of CB vevs, reducing the computational burden of the calculations.

The analysis of the dark matter phase of the HTM also shows interesting features: deeper CB vacua are possible
for well-defined regions of parameter space. We have deduced a necessary and sufficient condition for
{\em absolute} stability of the $N2$ minimum in the HTM potential with intact global symmetry in
eq.~\eqref{eq:bou}: it is required that
\be
M^2\,>\,-\,\sqrt{\mbox{min}\left(\lambda_2 + \dfrac{1}{2} \lambda_3\,,\,\lambda_2 + \lambda_3\right)}
\,\frac{m_h\,v}{\sqrt{2}}\,.
\ee
This is an extremely simple condition to include in one's parameter scan, and was possible to obtain
from the analytical expressions we deduced relating the depths of the potential at $CB$ and neutral vacua. Of course
this condition does not include the possibility of an $N2$ minimum being metastable but with a large enough lifetime
(larger than the age of the universe). That possibility would require a detailed calculation of tunneling times
to the deeper CB minimum, with a likely loosening of the above condition -- a similar situation occurs in
the 2HDM, where absolute stability conditions for neutral minima are loosened if tunneling times are taken into
account~\cite{Branchina:2018qlf}. That is certainly an interesting question to address, but it is outside the scope of the present work.

As for the softly broken model, the fact that a generic parameter space scan
was seen to have, for about 48\% of all combinations of parameters found, deeper CB vacua is troubling --
all the more so because this parameter scan, albeit not an exhaustive one,
yielded nevertheless phenomenologically acceptable (and interesting) scalar masses on a minimum of type $N1$,
where both triplet and doublet acquired vevs. Unlike the dark matter case, the CB vacua do not seem to concentrate in a well-defined region of parameter space, so the message to take from our results seems to be that any choice of
parameters that one wishes to study ought to be checked for the possibility of deeper CB vacua. We have
identified many possible CB vacua (both with real or complex vevs) which may be deeper than a $N1$ minimum -- in fact
only a few $CB$ vacua, such as $CB4$, are guaranteed to not be deeper than a $N1$ minimum. Unfortunately, short
of a numerical minimization for each set of potential parameters which yields $N1$ minima to check whether
it is the global one, there does not seem to be an analytical way to enquire about the stability of the potential.

To conclude, the analytical method used to compute the difference in the depth of the scalar potential at
different extrema -- neutral and charge breaking ones -- was incredibly useful and allowed the identification
of numerous possibilities of charge breaking vacua being possible, and even the obtention of bounds on the model's
parameters to avoid them, in some cases. However, we must remember that the method employed here is a tree-level
analysis only. For the 2HDM, the inclusion of 1-loop corrections, using the effective formalism approach, has been
shown to be able to change the stability picture deduced at tree-level~\cite{Ferreira:2015pfi,Ferreira:2019bij}.
Almost certainly the same will happen with the HTM, and one can expect, for instance, that the bound of
eq.~\eqref{eq:bou} will be relaxed once loop corrections will be taken into account. However, if the previous
work in the 2HDM has taught us anything regarding loop effects in the stability picture of the model, it is that
eventual changes to tree-level expectations exist but are rare, being confined to specific regions of parameter
space. Thus the tree-level analysis of the current work can be relied upon with confidence to provide good
guidance to the occurrence of charge breaking minima in the Higgs Triplet Model.

\subsubsection*{Acknowledgments}
This work began during a visit to the University of Toyama in Japan, whose hospitality PF is much
grateful for. PF sincerely acknowledges the insightful and interesting discussions during that period with
Shinya Kanemura and Hiroaki Sugiyama on the Higgs Triplet model, and Lu\'{\i}s Lavoura for pointing out
an error in the first version of the manuscript.
PF is supported in part by a CERN grant CERN/FIS-PAR/0002/2017, an FCT
grant PTDC/FIS-PAR/31000/2017, by the CFTC-UL strategic project
UID/FIS/00618/2019 and by the NSC, Poland, HARMONIA UMO-2015/18/M/ST2/00518. The work of BLG is supported by the FCT grant SFRH/BD/139165/2018. BLG dedicates his contribution to this work to the memory of his father.

\appendix

\section{Complex charge breaking vevs - non-soft breaking case}
\label{ap:nsbc}

Let us now consider complex entries in the $CB$ extrema. Considering that without the soft-breaking
term $\mu$ we have the freedom to independently rephase both the doublet and the triplet,
there are only three non-trivial different possibilities:
\ba
\langle \Phi\rangle_{CB2_{c}} = \frac{1}{\sqrt{2}}\,\left(
\begin{array}{c} 0 \\ c_1 \end{array}
\right)\;\;\; &,& \;\;\; \langle \Delta \rangle_{CB2_{c}} = \frac{1}{\sqrt{2}}\,
\left( \begin{array}{cc} 0 & c_3^{r} + i \, c_3^{i} \\
c_2 & 0 \end{array}  \right)
\\
\langle \Phi\rangle_{CB3_{c}} = \frac{1}{\sqrt{2}}\,\left(
\begin{array}{c} 0 \\ c_1 \end{array}
\right)\;\;\; &,& \;\;\; \langle \Delta \rangle_{CB3_{c}} = \frac{1}{\sqrt{2}}\,
\left( \begin{array}{cc} c_3/\sqrt{2} & c_4^{r} + i \, c_4^{i} \\
c_2 & -c_3/\sqrt{2} \end{array}  \right)
\\
\langle \Phi\rangle_{CB5_{c}} = \frac{1}{\sqrt{2}}\,\left(
\begin{array}{c} 0 \\ c_1 \end{array}
\right)\;\;\; &,& \;\;\; \langle \Delta \rangle_{CB5_{c}} = \frac{1}{\sqrt{2}}\,
\left( \begin{array}{cc} c_2/\sqrt{2} & c_3^{r}+ i \, c_3^i \\
0 & -c_2/\sqrt{2} \end{array}  \right)\,.
\ea
Following now the procedure explained in section~\ref{sec:nsb} for coexistence of these complex CB vevs
with an $N2$ extremum, we obtain:
\ba
V_{CB2_{c}} - V_{N2} &=& \frac{1}{4} \,\left[c_2^2 \, m^2_{H,A} \,+\,\left( {c_3^r}^2 \,+\, {c_3^i}^2 \right) \, m^2_{++}\right] \nonumber \\
& & \nonumber \\
V_{CB3_{c}} - V_{N2} &=& \frac{1}{4} \,\left(c_2^2 \,+\,c_3^2 \,+\,{c_4^r}^{2} \,+\, {c_4^i}^{2} \right) \, m^{2}_{++} \nonumber \\
& & \nonumber \\
V_{CB5_{c}} - V_{N2} &=& \frac{1}{4} \,\left[c_2^2 \, m^2_{+} \,+\,\left( {c_3^r}^2 \,+\, {c_3^i}^2 \right) \, m^2_{++}\right] \,,
\ea
where the squared masses above are computed at $N2$, see eqs.~\eqref{eq:mH2}--\eqref{eq:mchch2}.
Thus one concludes that, if $N2$ is a minimum all of its squared scalar masses will be positive and
one always obtains $V_{CBi_{c}} - V_{N2}\,>\,0$ -- no CB extrema deeper than a $N2$ minimum can occur
in the HTM.

Considering now the $N1$ case, we get:
\ba
V_{CB2_{c}} - V_{N1} &=& \frac{1}{4} \, \left( {c_3^r}^2 \,+\, {c_3^i}^2 \right) \, m_{++}^2 \nonumber \\
& & \nonumber \\
V_{CB3_{c}} - V_{N1} &=& 0 \nonumber \\
& & \nonumber \\
V_{CB5_{c}} - V_{N1} &=& \dfrac{c_1^2 \, m^{2}_{+}}{ 4 \left( 2+\dfrac{v_\Phi^2}{v_\Delta^2}\right) } \,+\, \dfrac{1}{8} \, c_2^2 \, m_{++}^{2} \,+\, \dfrac{ \left( {c_{3}^{r}}^2 \,+\, {c_{3}^{i}}^2 \right) \, m_{+}^{2}}{2 \left( 1+\dfrac{2 v_\Delta^2}{v_\Phi^2}\right) } \,,
\ea
and thus once again absolute stability of $N1$ minima against charge breaking is found to hold.

\section{Complex charge breaking vevs - soft breaking case}
\label{ap:sbc}

In order to complete the study of the soft-breaking case, one has also to consider the possibility of
complex vevs. The presence in the potential of the term with the $\mu$ coefficient forbids independent
rephasings of both the doublet and the triplet fields. We gain one more phase, in comparison with the
situation without soft breaking where the $\mu$-term is absent, gathering a total of two complex phases.
We choose to consider the possibility of having complex VEVs in the charged and double-charged entries of
the triplet. With this in mind, we can have the following possibilities:
\ba
\langle \Phi\rangle_{CB1_{c}} = \frac{1}{\sqrt{2}}\,\left(
\begin{array}{c} 0 \\ c_1 \end{array}
\right)\;\;\; &,& \;\;\; \langle \Delta \rangle_{CB1_{c}} = \frac{1}{\sqrt{2}}\,
\left( \begin{array}{cc} -\dfrac{c_3^{r} + i \, c_3^{i}}{\sqrt{2}} & 0 \\
c_2 & \dfrac{c_3^{r} + i \, c_3^{i}}{\sqrt{2}} \end{array}  \right)
\\
\langle \Phi\rangle_{CB2_{c}} = \frac{1}{\sqrt{2}}\,\left(
\begin{array}{c} 0 \\ c_1 \end{array}
\right)\;\;\; &,& \;\;\; \langle \Delta \rangle_{CB2_{c}} = \frac{1}{\sqrt{2}}\,
\left( \begin{array}{cc} 0 & c_3^{r} + i \, c_3^{i} \\
c_2 & 0 \end{array}  \right)
\\
\langle \Phi\rangle_{CB3_{c}} = \frac{1}{\sqrt{2}}\,\left(
\begin{array}{c} 0 \\ c_1 \end{array}
\right)\;\;\; &,& \;\;\; \langle \Delta \rangle_{CB3_{c}} = \frac{1}{\sqrt{2}}\,
\left( \begin{array}{cc} c_3/\sqrt{2} & c_4^{r}+ i \, c_4^i \\
c_2 & -c_3/\sqrt{2} \end{array}  \right)
\\
\langle \Phi\rangle_{CB3_{c'}} = \frac{1}{\sqrt{2}}\,\left(
\begin{array}{c} 0 \\ c_1 \end{array}
\right)\;\;\; &,& \;\;\; \langle \Delta \rangle_{CB3_{c'}} = \frac{1}{\sqrt{2}}\,
\left( \begin{array}{cc} \dfrac{c_3^{r} + i \, c_3^{i}}{\sqrt{2}} & c_4 \\
c_2 & -\dfrac{c_3^{r} + i \, c_3^{i}}{\sqrt{2}}  \end{array} \right)
\\
\langle \Phi\rangle_{CB3_{c''}} = \frac{1}{\sqrt{2}}\,\left(
\begin{array}{c} 0 \\ c_1 \end{array}
\right)\;\;\; &,& \;\;\; \langle \Delta \rangle_{CB3_{c''}} = \frac{1}{\sqrt{2}}\,
\left( \begin{array}{cc} \dfrac{c_3^{r} + i \, c_3^{i}}{\sqrt{2}} & c_4^{r} + i \, c_4^{i} \\
c_2 & -\dfrac{c_3^{r} + i \, c_3^{i}}{\sqrt{2}}  \end{array} \right)
\\
\langle \Phi\rangle_{CB4_{c}} = \frac{1}{\sqrt{2}}\,\left(
\begin{array}{c} 0 \\ c_1 \end{array}
\right)\;\;\; &,& \;\;\; \langle \Delta \rangle_{CB4_{c}} = \frac{1}{\sqrt{2}}\,
\left( \begin{array}{cc} \dfrac{c_2^{r} + i \, c_2^{i}}{\sqrt{2}} & 0 \\
0 & -\dfrac{c_2^{r} + i \, c_2^{i}}{\sqrt{2}}  \end{array} \right)
\\
\langle \Phi\rangle_{CB5_{c}} = \frac{1}{\sqrt{2}}\,\left(
\begin{array}{c} 0 \\ c_1 \end{array}
\right)\;\;\; &,& \;\;\; \langle \Delta \rangle_{CB5_{c}} = \frac{1}{\sqrt{2}}\,
\left( \begin{array}{cc} c_2/\sqrt{2} & c_3^{r} + i \, c_3^{i} \\
0 & -c_2/\sqrt{2}  \end{array} \right)
\\
\langle \Phi\rangle_{CB5_{c'}} = \frac{1}{\sqrt{2}}\,\left(
\begin{array}{c} 0 \\ c_1 \end{array}
\right)\;\;\; &,& \;\;\; \langle \Delta \rangle_{CB5_{c'}} = \frac{1}{\sqrt{2}}\,
\left( \begin{array}{cc} \dfrac{c_2^{r} + i \, c_2^{i}}{\sqrt{2}} & c_3 \\
0 & -\dfrac{c_2^{r} + i \, c_2^{i}}{\sqrt{2}}  \end{array} \right)
\\
\langle \Phi\rangle_{CB5_{c''}} = \frac{1}{\sqrt{2}}\,\left(
\begin{array}{c} 0 \\ c_1 \end{array}
\right)\;\;\; &,& \;\;\; \langle \Delta \rangle_{CB5_{c''}} = \frac{1}{\sqrt{2}}\,
\left( \begin{array}{cc} \dfrac{c_2^{r} + i \, c_2^{i}}{\sqrt{2}} & c_3^{r} + i \, c_3^{i} \\
0 & -\dfrac{c_2^{r} + i \, c_2^{i}}{\sqrt{2}}  \end{array} \right)
\\
\langle \Phi\rangle_{CB6_{c}} = \frac{1}{\sqrt{2}}\,\left(
\begin{array}{c} 0 \\ c_1 \end{array}
\right)\;\;\; &,& \;\;\; \langle \Delta \rangle_{CB6_{c}} = \frac{1}{\sqrt{2}}\,
\left( \begin{array}{cc} 0 & c_2^{r} + i \, c_2^{i} \\
0 & 0 \end{array} \right)
\ea
It is possible that some of these vev configurations may be reduced to others via doublet-triplet gauge
transformations, but we will err on the side of fastidiousness and consider them all.

Then, performing the same sort of calculations shown in section~\ref{sec:sb}, it is possible to obtain expressions
for the potential depth differences between the CB and N1 extrema when complex vevs are allowed. Namely:
\ba
V_{CB1_{c}} - V_{N1} &=& \frac{1}{4} \, \dfrac{m_{+}^2 }{1+\dfrac{2v_{\Delta}^{2}}{v_{\Phi}^{2}}} \, \left( {c_3^r}^2 \,+\, {c_3^i}^2 \right)  + \dfrac{m_{A}^{2}}{4\left( 1 + \dfrac{4v_{\Delta}^{2}}{v_{\Phi}^{2}} \right) }\left( c_{2} - v_{\Delta} \right)^{2} \left( 1 - \dfrac{c_{1}^{2}}{v_{\Phi}^{2}} \dfrac{v_{\Delta}}{c_{2}} \right)    \nonumber \\
& & \nonumber \\
V_{CB2_{c}} - V_{N1} &=& \dfrac{m_{A}^{2}}{4\left( 1 + \dfrac{4v_{\Delta}^{2}}{v_{\Phi}^{2}} \right) }\left( c_{2} - v_{\Delta} \right)^{2} \left( 1 - \dfrac{c_{1}^{2}}{v_{\Phi}^{2}} \dfrac{v_{\Delta}}{c_{2}} \right) + \frac{1}{4} \, m_{++}^2 \, \left( {c_3^r}^2 \,+\, {c_3^i}^2 \right)   \nonumber \\
& & \nonumber \\
V_{CB4_{c}} - V_{N1} &=& \dfrac{m^{2}_{A}}{ 4 \left( 1+\dfrac{4v_\Delta^2}{v_\Phi^2}\right) } \left[ v_{\Delta}^{2} + c_{1}^{2}\dfrac{v_{\Delta}^{2}}{v_{\Phi}^{2}} + \dfrac{1}{2} \left( {c_2^r}^2 + {c_2^i}^2 \right) \right]  \,+\, \dfrac{1}{8} \, \left( {c_2^r}^2 + {c_2^i}^2 \right) \, m_{++}^{2} \,+\, \dfrac{v_{\Delta}^2}{v_{\Phi}^2} \dfrac{c_{1}^{2}m_{+}^{2}}{4\left( 1 + \dfrac{2v_{\Delta}^2}{v_{\Phi}^2}   \right) }  \nonumber \\
& & \nonumber \\
V_{CB5_{c}} - V_{N1} &=& \dfrac{m^{2}_{A}}{ 4 \left( 1+\dfrac{4v_\Delta^2}{v_\Phi^2}\right) } \left[ v_{\Delta}^{2} + c_{1}^{2}\dfrac{v_{\Delta}^{2}}{v_{\Phi}^{2}}   + \dfrac{1}{2} c_2^2 - \left( {c_3^r}^2 + {c_3^i}^2 \right) \right] \,+\, \dfrac{1}{8} \, {c_2}^2 \, m_{++}^{2} \nonumber \\
& & \nonumber \\
\,&+&\, \dfrac{m_{+}^{2}}{ 1 + \dfrac{2v_{\Delta}^2}{v_{\Phi}^2}}  \left[ \dfrac{1}{2} \left( {c_3^r}^2 + {c_3^i}^2 \right) + \dfrac{1}{4} c_1^2\dfrac{v_{\Delta}^2}{v_{\Phi}^2}  \right]  \nonumber \\
& & \nonumber \\
V_{CB5_{c'}} - V_{N1} &=& \dfrac{m^{2}_{A}}{ 4 \left( 1+\dfrac{4v_\Delta^2}{v_\Phi^2}\right) } \left[ v_{\Delta}^{2} + c_{1}^{2}\dfrac{v_{\Delta}^{2}}{v_{\Phi}^{2}}   + \dfrac{1}{2} \left( {c_2^r}^2 + {c_2^i}^2 \right) - {c_3}^2 \right] \,+\, \dfrac{1}{8} \, \left( {c_2^r}^2 + {c_2^i}^2 \right)  \, m_{++}^{2} \nonumber \\
& & \nonumber \\
\,&+&\, \dfrac{m_{+}^{2}}{ 1 + \dfrac{2v_{\Delta}^2}{v_{\Phi}^2}}  \left( \dfrac{1}{2} {c_3}^2 + \dfrac{1}{4} c_1^2\dfrac{v_{\Delta}^2}{v_{\Phi}^2}  \right)  \nonumber \\
& & \nonumber \\
V_{CB5_{c''}} - V_{N1} &=& \dfrac{m^{2}_{A}}{ 4 \left( 1+\dfrac{4v_\Delta^2}{v_\Phi^2}\right) } \left[ v_{\Delta}^{2} + c_{1}^{2}\dfrac{v_{\Delta}^{2}}{v_{\Phi}^{2}}   + \dfrac{1}{2} \left( {c_2^r}^2 + {c_2^i}^2 \right) - {c_3^r}^2 - {c_3^i}^2 \right] \,+\, \dfrac{1}{8} \, \left( {c_2^r}^2 + {c_2^i}^2 \right)  \, m_{++}^{2} \nonumber \\
& & \nonumber \\
\,&+&\, \dfrac{m_{+}^{2}}{ 1 + \dfrac{2v_{\Delta}^2}{v_{\Phi}^2}}  \left[ \dfrac{1}{2} \left( {c_3^r}^2 + {c_3^i}^2 \right)  + \dfrac{1}{4} c_1^2\dfrac{v_{\Delta}^2}{v_{\Phi}^2}  \right]  \nonumber \\
& & \nonumber \\
V_{CB6_{c}} - V_{N1} &=& \dfrac{m^{2}_{A}}{ 4 \left( 1+\dfrac{4v_\Delta^2}{v_\Phi^2}\right) }
\left( v_{\Delta}^{2} - {c_2^r}^2 - {c_2^i}^2 \right)  \,+\,  \dfrac{m_{+}^{2}}{2 \left(  1 +
\dfrac{2v_{\Delta}^2}{v_{\Phi}^2} \right) }  \left( {c_2^r}^2 + {c_2^i}^2 +
c_1^2\dfrac{v_{\Delta}^2}{v_{\Phi}^2}  \right)
\ea
When considering one of the three vev configurations $CB3_{c}$, $CB3_{c'}$ or $CB3_{c''}$ simultaneously
with the $N1$ state, it is not possible to find a solution for the minimisation conditions. Which means that
in fact the $CB3_{c}$, $CB3_{c'}$ and $CB3_{c''}$ extrema do not exist simultaneously with a $N1$ vaccum.

The conclusion to draw from these lengthy expressions is that only the $CB4_{c}$ case guarantees a potential
difference positive when $N1$ is a minimum. For all the others, an $N1$ minimum is not guaranteed to be
deeper than a CB extremum, unlike the $\mu = 0$ case.

\section{Multiple $N1$ minima}
\label{ap:n1l}

The minimisation conditions for an $N1$-type minimum admit more than one solution. Other than trivial
sign changes for the vevs, we can obtain  different values for the doublet and triplet vevs, corresponding
to minima which break the same symmetries but nonetheless yield different physics -- different numerical
values for the doublet vev, for instance, would originate quarks with masses different from the known
ones. The electroweak gauge boson masses would also change. An analogous situation occurs within the 2HDM,
originating the so-called ``panic
vacua"\cite{Ivanov:2006yq,Ivanov:2007de,Barroso:2007rr,Barroso:2012mj,Barroso:2013awa}. Within the HTM,
the minimisation conditions which determine the vevs of an $N1$ extremum (defined in eq.~\eqref{eq:N1})
are
\ba
\frac{\partial V}{\partial v_\Phi} &=& \,v_\Phi\,\left[ m^2 \,-\,\sqrt{2} \mu v_\Delta \,+\,
\lambda_1 v_\Phi^2 \,+\, \frac{1}{2} (\lambda_4 + \lambda_5) v_\Delta^2\right]\,=\,0\,,
\nonumber \\
\frac{\partial V}{\partial v_\Delta} &=& \,M^2 v_\Delta\,-\,\frac{\mu}{\sqrt{2}}\,v_\Phi^2
\,+\,(\lambda_2 + \lambda_3) v_\Delta^3\,+\, \frac{1}{2} (\lambda_4 + \lambda_5)\,
 v_\Phi^2 v_\Delta\,.
\ea
One can eliminate $v_\Phi$ from the first equation and replace it in the second one, obtaining
a cubic equation for $v_\Delta$, namely
\be
a_3\,v_\Delta^3 \,+\,a_2\,v_\Delta^2\,+\,a_1\,v_\Delta\,+\,a_0\,=\,0\,,
\label{eq:v2c}
\ee
with
\ba
a_0 &=& 4\,\mu\,m^2 \nonumber\\
a_1 &=& -\,2\sqrt{2} \left[(\lambda_4 + \lambda_5)\,m^2 - 2\lambda_1 M^2 + 2 \mu^2 \right]\nonumber\\
a_2 &=& 6\,(\lambda_4 + \lambda_5)\,\mu \nonumber\\
a_3 &=& \sqrt{2} \left[4\lambda_1 (\lambda_2 + \lambda_3) - (\lambda_4 + \lambda_5)^2\right]\,.
\ea
Thus we have the possibility of multiple minima of type $N1$ -- notice that this is only possible
if $\mu\neq 0$, otherwise $a_0 = 0$ and eq.~\eqref{eq:v2c} has only two solutions related by a
minus sign (and therefore a single minimum). For the HTM with a softly broken global symmetry there
is then the possibility of a solution ($N1$) where the doublet
and triplet have vevs such that $v_\Phi^2 + 2v^2_\Delta \simeq$(246 GeV)$^2$; and other minima ($N1^\prime$),
 with different vevs $v_\Phi^\prime$ and $v^\prime_\Delta$, for which $v_\Phi^2 + 2v^2_\Delta \neq$(246
  GeV)$^2$
-- which would originate electroweak breaking, but with a completely different mass spectrum for
gauge bosons and fermions, and thus forbidden by experimental evidence. Parameter combinations which would
originate deeper vacua of this type should therefore be excluded. Using a bilinear calculation, it is possible
to express the relative depth of the potential of two extrema of types $N1$ and $N1^\prime$, to wit
\be
V_{N1^\prime}\,-\,V_{N1} \,=\,\frac{1}{4}\,\left[\frac{m^2_A}{1 + 4\left(\displaystyle{
\frac{v_\Delta}{v_\Phi}}\right)^2}\,-\,
\frac{{m^2_A}^\prime}{1 + 4\left(\displaystyle{\frac{v^\prime_\Delta}{v^\prime_\Phi}}\right)^2}\right]\,
\left(v_\Delta \,-\, v^\prime_\Delta\right)^2\, ,
\ee
where $m_A$ ($m^\prime_A$) is the pseudoscalar mass at the $N1$ ($N1^\prime$) extremum. As we see,
an $N1$ minimum is not guaranteed to be stable, it could coexist with a deeper $N1^\prime$ minimum.
We have verified, for the whole of the parameter space we scanned, that the $N1$ minima we found
have no deeper $N1^\prime$ extrema. This verification is quite simple, since the new vevs at $N1^\prime$ would
be the additional roots of the cubic equation~\eqref{eq:v2c}. That we have found no combination of parameters
for which deeper $N1^\prime$  minima do not exist does not mean that such coexistence
is impossible, simply that it should be vary rare to find parameter combinations that allow for
it -- that, at least, is what one
would expect from an analogy with the similar ``panic vacuum" situation in the 2HDM.

\bibliography{references}

\end{document}